\documentclass[preprint,showpacs,preprintnumbers,amsmath,amssymb]{revtex4}

\usepackage{graphicx}
\usepackage{dcolumn}
\usepackage{bm}
\usepackage{relsize}
\usepackage{mdwlist}
\usepackage{soul}
\usepackage{hyperref}
\usepackage[all]{hypcap}

\def\cesrta{{C{\smaller[2]ESR}TA}}

\begin{document}



\title{Beam Induced Electron Cloud Resonances in Dipole Magnetic Fields}

\author{J.~R.~Calvey\footnote{Present address: Advanced Photon Source, Argonne National Laboratory, Argonne, IL}}
\author{W.~Hartung\footnote{Present address: Facility for Rare Isotope Beams, Michigan State
University, East Lansing MI}}
\author{J.~Makita\footnote{Present address: Department of Physics, Old Dominion University, Norfolk, VA}}
\affiliation{Cornell Laboratory for Accelerator-based Sciences and
  Education, Cornell University, Ithaca, NY}

\author{M.~Venturini}
\affiliation{Lawrence Berkeley National Laboratory, Berkeley, CA}

\date{\today}

\begin{abstract}
        The buildup of low energy electrons in an accelerator, known as electron cloud, can be severely detrimental to machine performance.  Under certain beam conditions, the beam can become resonant with the cloud dynamics, accelerating the buildup of electrons.  This paper will examine two such effects: multipacting resonances, in which the cloud development time is resonant with the bunch spacing, and cyclotron resonances, in which the cyclotron period of electrons in a magnetic field is a multiple of bunch spacing.  Both resonances have been studied directly in dipole fields using retarding field analyzers installed in the Cornell Electron Storage Ring (CESR).  These measurements are supported by both analytical models and computer simulations.


\end{abstract}

\maketitle

\section{\label{sec:intro} Introduction}

As a part of the \cesrta~ program at Cornell~\cite{1748-0221-10-07-P07012}, the Cornell Electron Storage Ring (CESR) was instrumented with several retarding field analyzers (RFAs)~\cite{NIMA453:507to513}, to study the buildup of low energy electrons in an accelerator vacuum chamber.  This effect, known as electron cloud~\cite{ECLOUD12:Miguel,doi:10.1142/S0217751X14300233}, has been observed in a number of machines~\cite{PRSTAB7:024402,PRSTAB14:071001,NIMA556:399to409,PRSTAB6:034402,PAC09:WE4GRC02,PhysRevSTAB.6.014203,PRSTAB16:011003}, and is known to cause emittance growth and beam instabilities~\cite{PhysRevSTAB.7.124801}.  It is especially dangerous for low emittance, positively charged beams, and is expected to be a limiting factor in next generation positron and proton storage rings, such as the International Linear Collider damping ring~\cite{ILCREP2007:001,PRSTAB17:031002}.

In lepton machines, electron cloud is usually seeded by photoelectrons generated by synchrotron radiation.  The collision of these electrons with the beam pipe can then produce one or more secondary electrons, depending on the secondary electron yield (SEY) of the material.  The SEY depends on the energy and angle of the incident electron~\cite{PRSTAB5:124404}, with peak secondary production occurring at $E_{max} \approx 300$~eV.  If the average SEY is greater than unity, the cloud density will grow exponentially, until a saturation is reached.  Most secondary electrons are generated with low energy ($<$ 10~eV), but can be given additional energy by the beam.  As we will show in this paper, an unfortunate choice of beam parameters (particulary bunch spacing and charge) can drive up the average electron energy up into a regime of high secondary production (near $E_{max}$), resulting in a higher cloud density.

Retarding field analyzers provide information on the local electron cloud density, energy, and transverse distributions.  Previous papers have described the use of RFAs at \cesrta~ to directly compare different electron cloud mitigation techniques~\cite{NIMA760:86to97,NIMA770:141to154}.  In addition, computer simulations have been compared to RFA measurements, to quantify the electron emission properties of different cloud mitigating coatings in field free regions~\cite{PRSTAB17:061001}.  Simulations of cloud dynamics in dipole and wiggler fields have been presented in conference proceedings~\cite{PAC09:FR5RFP043,IPAC10:TUPD022,IPAC11:MOPS083,IPAC12:WEPPR088}.  This paper will summarize and expand on these results.  In particular, multipacting and cyclotron resonances will be examined in detail.  These effects, in which resonant interactions between the beam and electrons lead to accelerated cloud development, should be avoided to ensure optimal machine performance.


\subsection{Retarding Field Analyzers}

A retarding field analyzer consists of three main components~\cite{NIMA453:507to513}: holes drilled in the beam pipe to allow electrons to enter the device; a retarding grid, to which a voltage can be applied, rejecting electrons with less than a certain energy; and a positively biased collector, to capture any electrons which make it past the grid.  If space permits, additional (grounded) grids can be added to produce a more ideal retarding field.  In addition, the collectors of most RFAs used in \cesrta~are segmented to allow characterization of the spatial structure of the cloud build-up.  Thus a single RFA measurement provides information on the local cloud density, energy, and transverse distribution.  Some of the data presented here are voltage scans, in which the retarding voltage is varied (typically from +100 to $-250$~V or $-400$~V) while beam conditions are held constant.  In other measurements, where we want to study the detector response as a function of some external parameter (e.g. bunch spacing), the retarding grid was biased at +50~V, to capture all incoming electrons.  The collector was set to +100~V for all of our measurements.

\begin{figure}
\centering
\includegraphics[width=.6\textwidth]{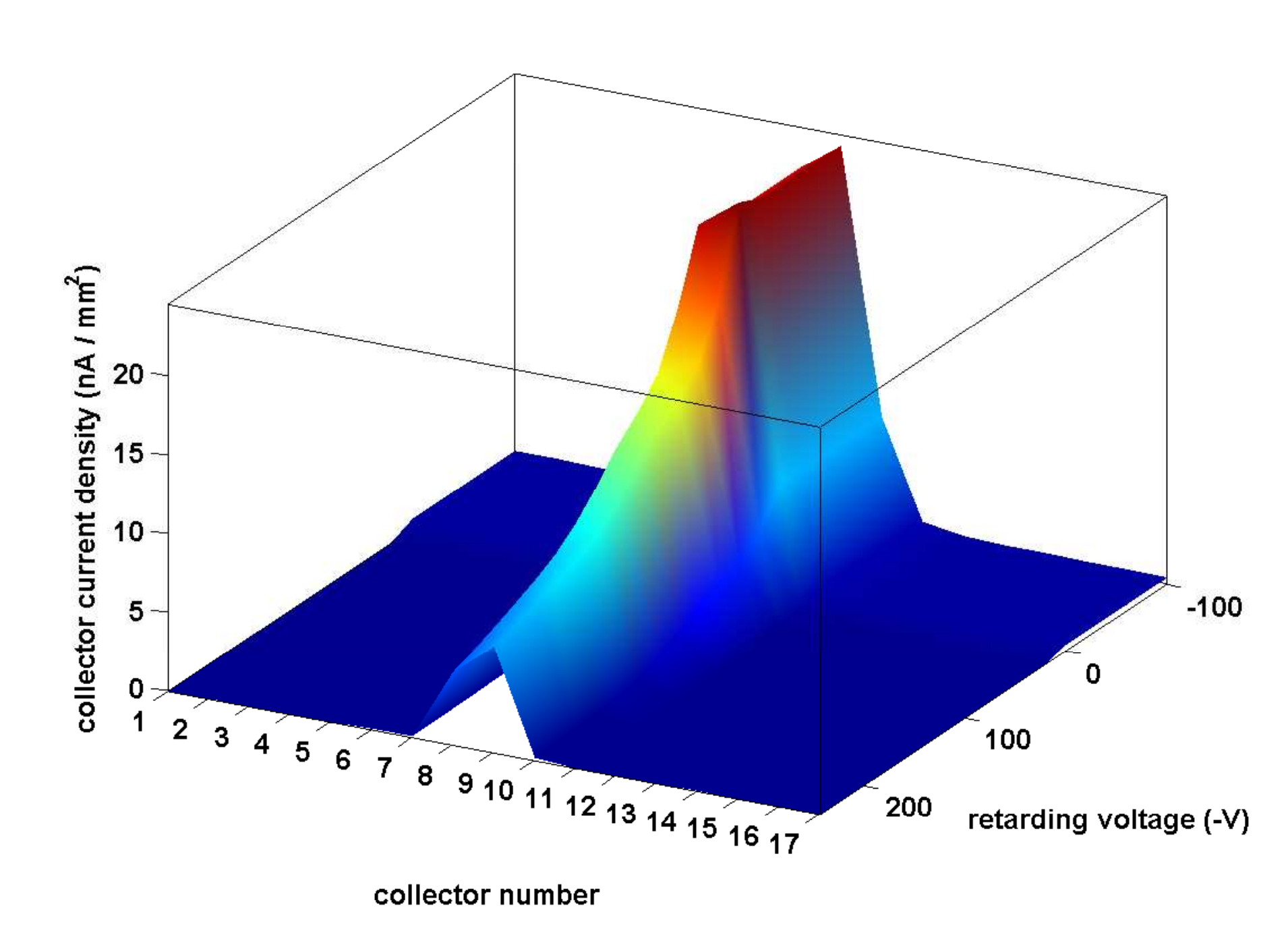} \\    
\caption[CESR dipole RFA voltage scans]{\label{fig:chic_dipole_meas} Dipole RFA voltage scan: 1x45x1.25~mA $e^+$, 14~ns, 5.3~GeV, 810~gauss field.  The central collector is no. 9.}
\end{figure}

An example voltage scan is given in Fig.~\ref{fig:chic_dipole_meas}.  The RFA response is plotted as a function of collector number and retarding voltage.  Roughly speaking, this is a description of the transverse and energy distribution of the cloud.  Collector  1 is closest to the outside of the chamber (where direct synchrotron radiation hits).  The signal is strongly peaked in the central collector (no. 9), which is aligned with the horizontal position of the beam.  The sign convention for retarding voltage is chosen so that a positive value on this axis corresponds to a negative physical voltage on the grid (and thus a rejection of lower energy electrons).  The beam conditions are given as ``1x45x1.25~mA e$^+$, 14~ns, 5.3~GeV."  This notation indicates one train of 45 positron bunches, with a per-bunch current of 1.25~mA (1~mA = $1.6\times10^{10}$ particles), with 14~ns bunch spacing, and a beam energy of 5.3~GeV.


\subsection{Electron Cloud in Dipoles}

In the presence of a dipole magnetic field, an electron will undergo helical motion, spiralling around the field lines.  For a standard dipole magnet in an accelerator (with strength $\sim$ 1~kilogauss), a typical cloud electron (with energy $\sim$ 10 - 100~eV) will have a cyclotron radius on the order of a few hundred $\mu$m.  In other words, the motion of the electron will be approximately one dimensional, along the direction of the dipole field.  This pinning of the motion to the field lines often results in a strong concentration of the cloud in the center of the chamber, where beam kicks are strongest.  Stronger beam kicks drive the average electron energy up, which typically results in a higher average SEY (since most secondary electrons are emitted with $E_{sec} \ll E_{max}$).  This effect is seen clearly in Fig.~\ref{fig:chic_dipole_meas}.  In addition, multipacting and cyclotron resonances, described below, can appear in dipole fields.



\subsubsection{\label{ssec:mult} Multipacting Resonances}

 A multipacting resonance occurs when a characteristic time for the cloud development is equal to the bunch spacing.  As originally proposed by Gr\"{o}bner~\cite{HEACC77:GROBNER}, this happens when the kick from the beam gives secondary electrons near the vacuum chamber wall just enough energy to reach the opposite wall in time for the next bunch.  These electrons generate more secondaries, which are again given energy by the beam.  This process continues, resulting in a resonant buildup of the cloud.  The resonant condition is given by Eq.~(\ref{eq:grob}).


\begin{equation}
    \label{eq:grob}
    t_b = \frac{b^2}{c r_e N_b}
\end{equation}


Here $t_b$ is the bunch spacing, $b$ is the chamber half-height, $c$ is the speed of light, $r_e$ is the classical electron radius, and $N_b$ is the bunch population.  A more general condition was derived by Harkay et al.~\cite{PAC03:RPPG002,PRSTAB6:034402}, which includes nonzero secondary emission velocity.  In Section~\ref{ssec:mult_res}, we develop an even more general model of multipacting resonances, which includes the possibility of multiple beam kicks.


\subsubsection{\label{ssec:cyc} Cyclotron Resonances}

A cyclotron resonance occurs when the bunch spacing is an integral multiple of the cyclotron period of an electron in a dipole field~\cite{PRSTAB11:091002}.  Under these conditions, the transverse beam kick to a given electron will always be in the same direction, resulting in a steady increase in the particle's energy, and (usually) a higher secondary electron yield when it hits the vacuum chamber wall.  The resonant condition is given in Eq.~(\ref{eq:cyc}), where $m_e$ is the electron mass, $q_e$ is the electron charge, $n$ is an integer, and $B$ is the magnetic field strength.


\begin{equation}
    \label{eq:cyc}
    t_b = \frac{2 \pi m_e n}{q_e B}
\end{equation}

Cyclotron resonances were observed at SLAC using a chicane of four dipole magnets instrumented with RFAs~\cite{NIMA621:33to38}.  Unexpectedly, the resonances sometimes appeared as peaks in the signal, and other times as dips.  This chicane was moved to CESR early in the \cesrta~program.  In Section~\ref{ssec:cyc_res}, we confirm the existence of cyclotron resonances, and in Section~\ref{ssec:cyc_res_sim}, we provide an explanation for the peak/dip phenomenon.

\section{\label{sec:instrumentation} Instrumentation}

Detailed descriptions of the \cesrta~electron cloud experimental program, design of the field region RFAs, and data acquisition system can be found elsewhere~\cite{NIMA770:141to154,CLNS:12:2084}; here we provide only a brief summary.  RFAs in each field region had to be specially designed to fit inside the narrow magnet apertures.  The key parameters of each RFA type are listed in Table~\ref{tab:dipole_rfa_styles}.


\begin{table}
  \centering
  \footnotesize
     \setlength{\tabcolsep}{10pt}
    \caption{\label{tab:dipole_rfa_styles} List of dipole/wiggler RFA locations.  The elliptical and rectangular chambers are 9~cm in width by 5~cm in height.  The circular chamber is 4.5~cm in radius.  ``Grid trans." refers to the optical transparency of the grids.  Note that the wiggler RFAs used two generations of grids with different transparencies.}
  \begin{tabular}{ccccccc}
    \hline \hline
  RFA  &   Chamber type   &     Field Strength &   Grids  &   Collectors  & Grid trans. \\
  \hline
  CESR dipole   &   Elliptical Al   & 0.079 - 0.2010~T    &  1   &   9 &   ~38\% \\
  Chicane dipole   &   Circular Al  &  0 - 0.12~T       &  3   &   17 &   ~92\% \\
  Wiggler       &   Rectangular Cu  &   1.9~T           &  1   &   12 &   ~38/92\% \\
  \hline \hline
  \end{tabular}
\end{table}

\paragraph{CESR Dipole RFA} To study cloud buildup in a realistic dipole field environment, a thin RFA was installed inside a CESR dipole magnet.  The magnetic field in this magnet depends on the beam energy: 790~gauss at 2.1~GeV, 1520~gauss at 4~GeV, and 2010~gauss at 5.3~GeV.  The chamber is made of uncoated (6063) aluminum.


\paragraph{Chicane RFAs} A chicane of four dipole magnets designed at SLAC~\cite{NIMA621:33to38} was installed in the L3 straight.  The field of these magnets can be varied over the range of 0 to 1.46~kilogauss, which allowed for the study of the effect of dipole field strength on cloud dynamics, without affecting the trajectory of stored beams in the rest of the ring.  Three of the chicane dipole chambers tested different electron cloud mitigation techniques: two of the chambers were TiN coated~\cite{NIMA551:187to199}, and one was both grooved~\cite{NIMA571:588to598,PAC07:THPMN118} and TiN coated (the fourth was bare aluminum).


\paragraph{Wiggler RFAs} During the \cesrta~reconfiguration in 2008, six superconducting wigglers were installed in the L0 straight section of CESR.  They were typically operated with a peak transverse field of 1.9~T.  Three of these wigglers were instrumented with RFAs, at three different locations in the wiggler field: in the center of the wiggler pole (effectively a 1.9~T dipole field), half way between two poles (where the field is longitudinal), and in an intermediate region~\cite{NIMA770:141to154}.  This paper will focus on the pole center RFAs.

The first generation wiggler RFAs were equipped with low-transparency stainless steel grids.  However, as described in Section~\ref{ssec:tramp}, secondary emission from these grids lead to a significant interaction between the electron cloud and the RFA, complicating the interpretation of the measurements.  Consequently, in the second generation of wiggler chambers, the grids were changed to high-transparency copper meshes.  The use of high transparency grids effectively solved the grid emission problem.


\section{Measurements and Analytical Models}

Many measurements have been taken in CESR with RFAs in dipole fields, under a wide variety of different beam conditions.  This has allowed for detailed studies of electron cloud dynamics, in particular of multipacting and cyclotron resonances.

\subsection{\label{ssec:mult_res} Multipacting Resonances}



\begin{figure}
   \centering
   \begin{tabular}{c}
   \includegraphics*[width=0.6\textwidth]{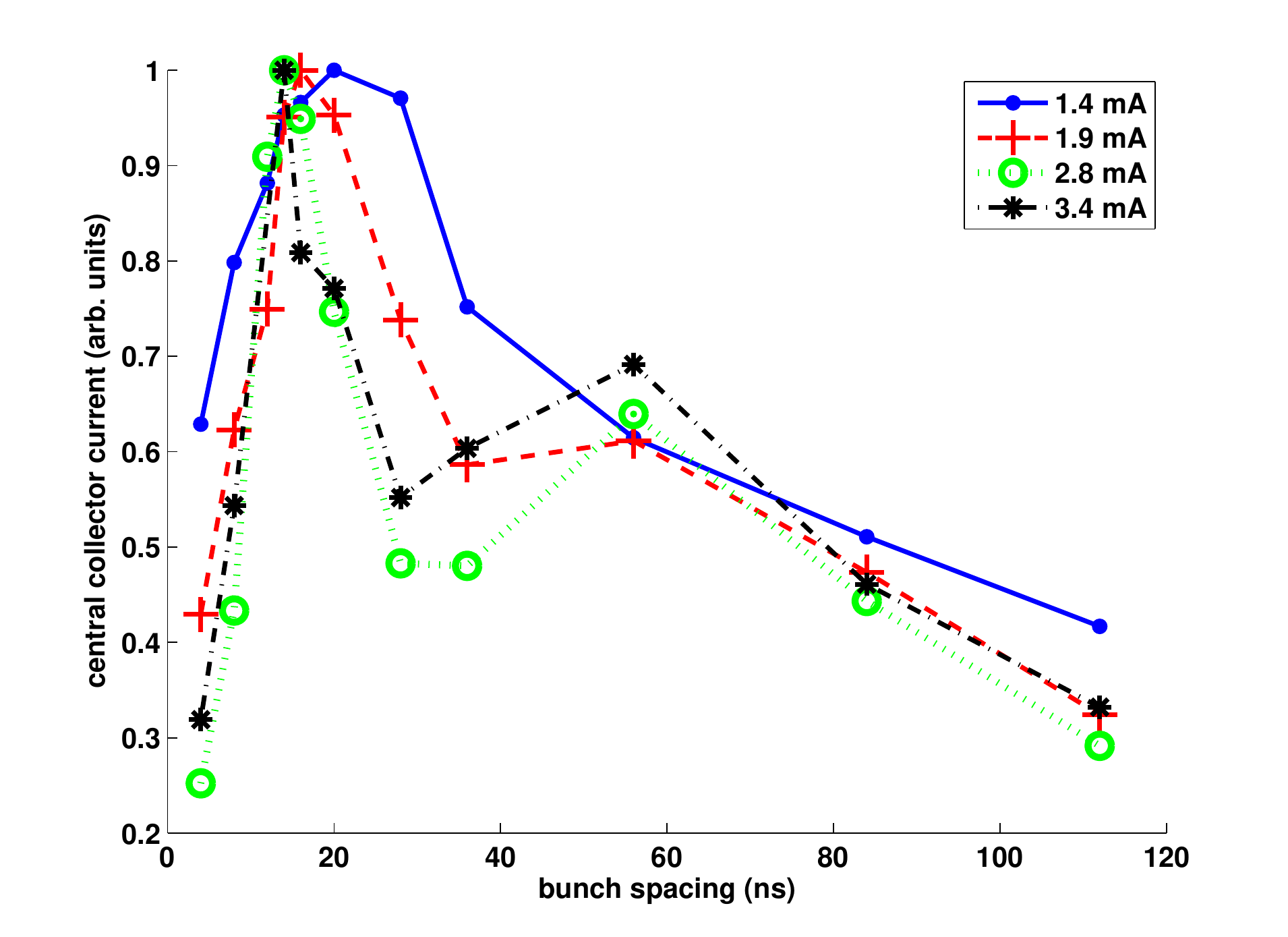} \\ 
   \includegraphics*[width=0.6\textwidth]{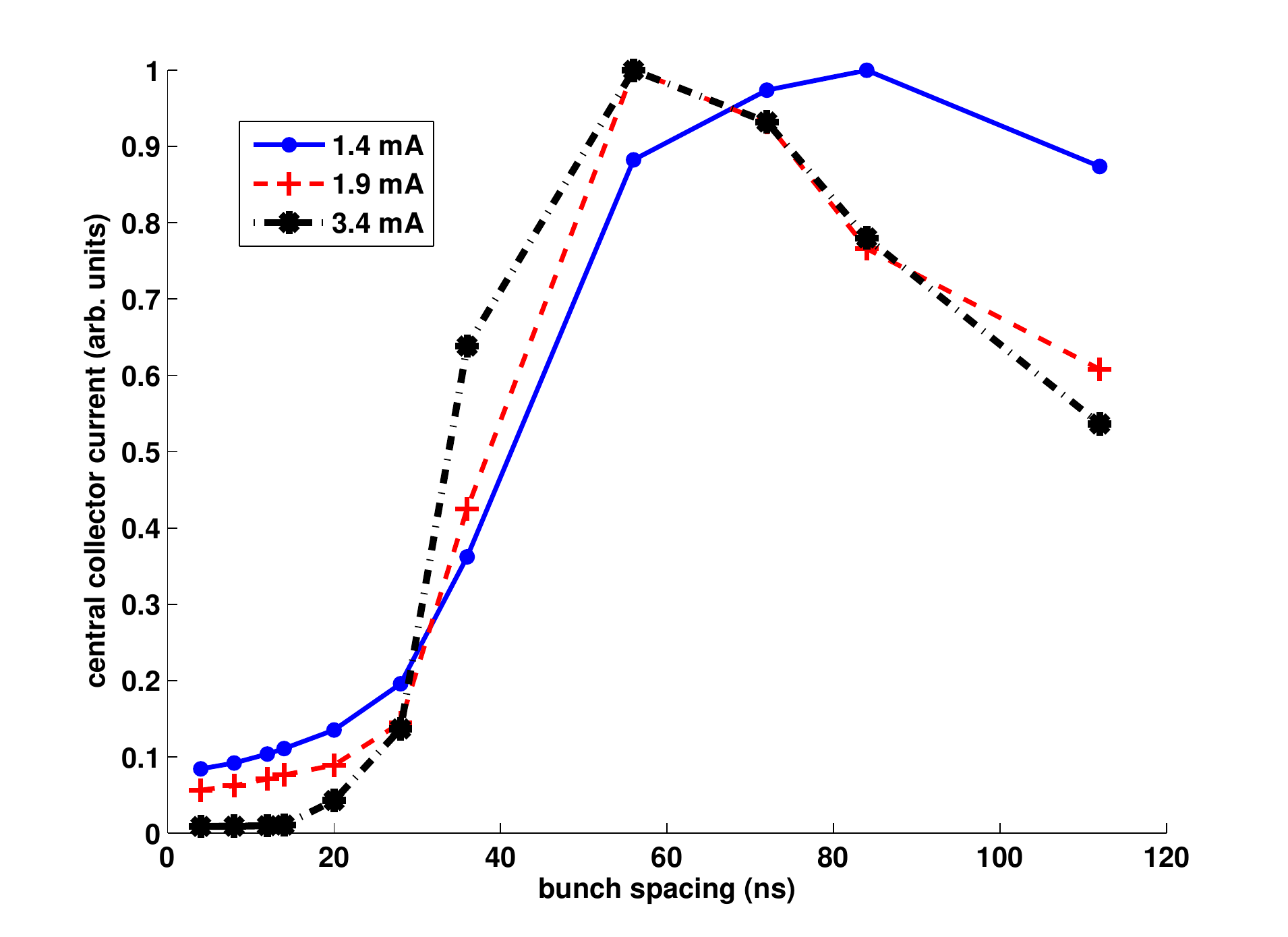} 
   \end{tabular}
\caption{\label{fig:dipole_spacing_chic} Central collector signal in the chicane dipole RFA (set to 810~gauss) as a function of bunch spacing, at different bunch currents. Top: positron beam; bottom: electron beam.  Note that the signals have been normalized to be on the same scale.  In absolute terms, the peak positron signal was about five times the peak electron signal.}
\end{figure}

To study the time evolution of the electron cloud, we collected RFA data with bunch spacings varying from 4~ns to 112~ns.  All of the data presented in this section were taken with a single train of 20 bunches, at beam energy 5.3~GeV.  Fig.~\ref{fig:dipole_spacing_chic} shows the signal in the central collector of the chicane RFA as a function of bunch spacing, for different bunch currents, and for both electron and positron beams.  A few interesting features are  readily apparent in the data. Except at the lowest current value, both the electron and positron beam data show a peak at 56 ns.  The positron data has another peak, which moves to lower bunch spacings at higher currents.  These data are not consistent with a simple multipacting resonance (Eq.~(\ref{eq:grob})), which would account for only one resonance in the positron measurement, and none in the electron measurement.  Additionally, the beam kicks at the wall are very small for this case (amounting to 13~eV for a 3.5~mA beam), and so are unlikely to drive electrons at the wall into a regime of high secondary production.

\begin{figure}
   \centering
   \begin{tabular}{c}
   \includegraphics*[width=0.6\textwidth]{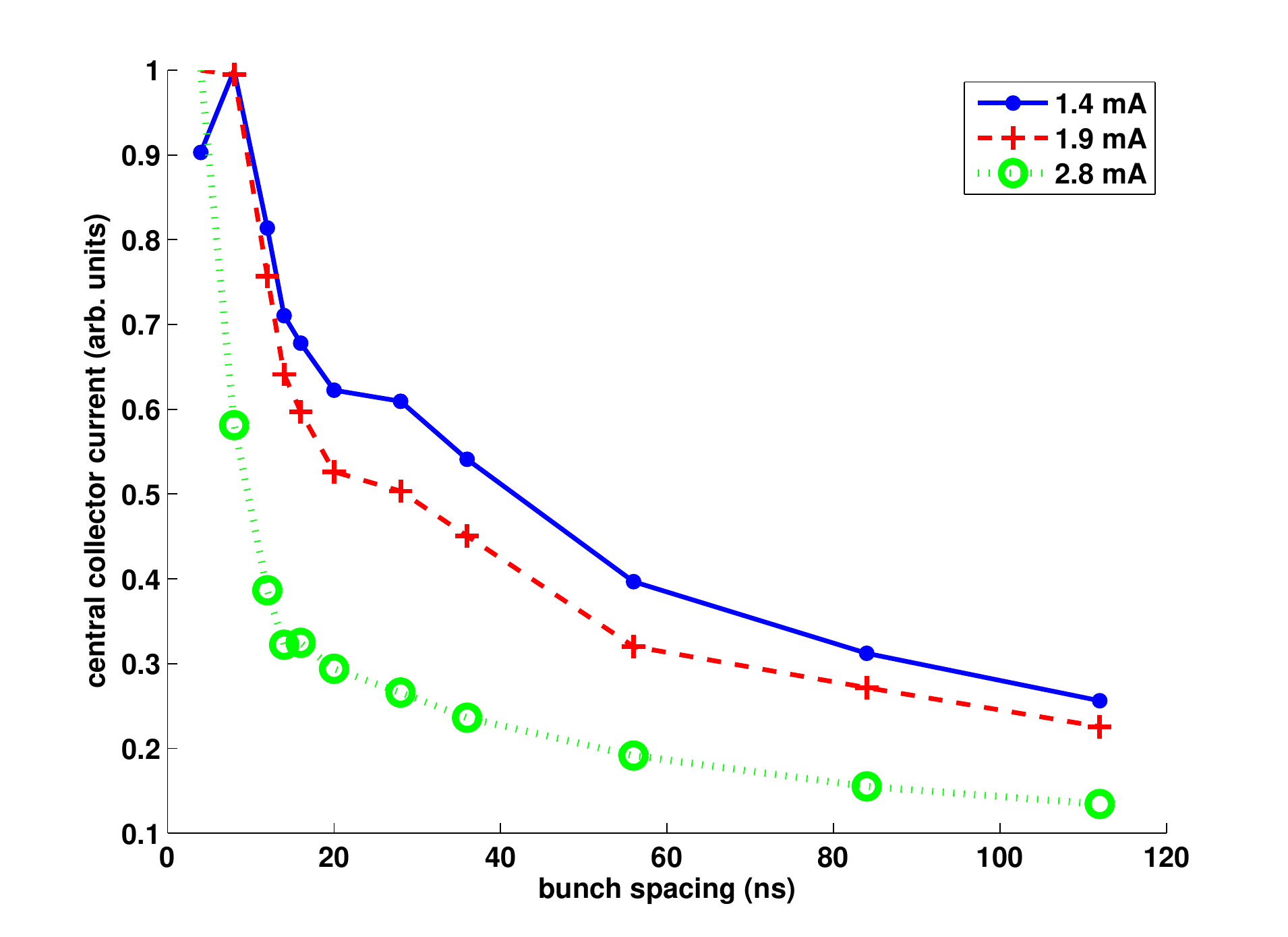} \\ 
   \includegraphics*[width=0.6\textwidth]{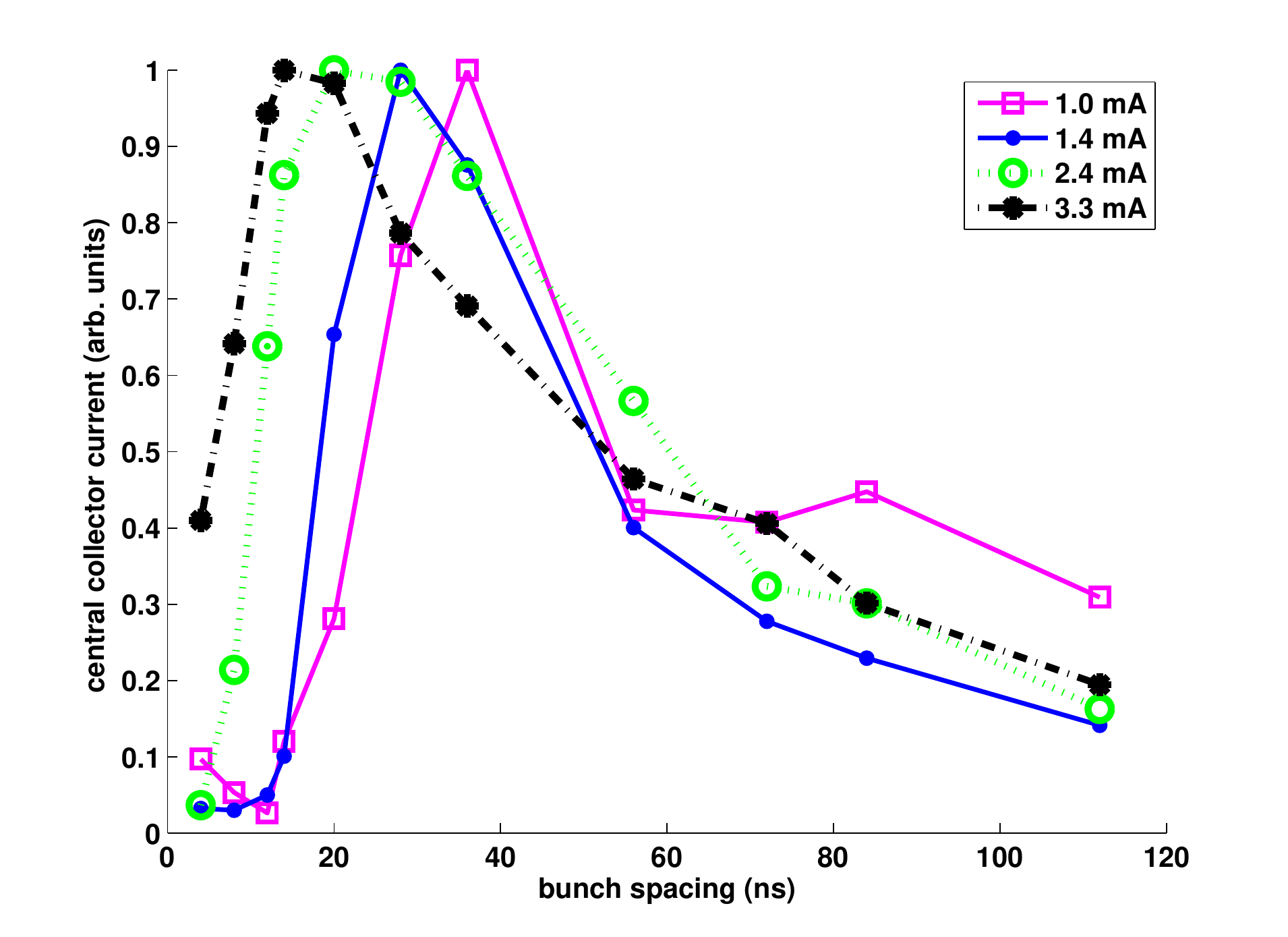} 
   \end{tabular}
\caption{\label{fig:dipole_spacing_cesr} Central collector signal in the CESR dipole RFA as a function of bunch spacing, at different bunch currents.  Top: positron beam; bottom: electron beam.  Note that the signals have been normalized to be on the same scale.  In absolute terms, the peak positron signal was about four times the peak electron signal.}
\end{figure}

A similar set of data for the CESR dipole RFA is shown in Fig.~\ref{fig:dipole_spacing_cesr}.  In this case, both the electron and positron beam data contain a single peak that moves to lower spacings as the current increases.  The positron data peaks occur at much lower spacings that the electron peaks.

\subsubsection{Analytical Model}

These resonances can be explained if we allow the secondary electrons to be generated with some (small) energy.  If the time for a typical secondary electron to travel to the center of the beam pipe is equal to the bunch spacing, this electron will be kicked strongly by the beam, and is likely to produce more secondary electrons~\cite{PRSTAB6:034402}.

If we ignore the time for the kicked electron to travel to the beam pipe wall, the resonance condition is given by Eq.~(\ref{eq:tb1_res}), where $t_b$ is the bunch spacing, $b$ is the chamber half-height (i.e. the distance from the wall to the beam), and $v_{sec}$ is a characteristic secondary electron velocity.


\begin{equation}
\label{eq:tb1_res}
t_b = b / v_{sec}
\end{equation}

For a (plausible~\cite{PRSTAB5:124404}) secondary emission energy of 1.5~eV, this peak will occur at 61~ns for the chicane dipole case ($b = 4.5$~cm).  Because aluminum has a high SEY for a broad range of incident energies, we expect the resonance to be somewhat broad.  The fact that there is a finite width to the secondary energy distribution will further smear out the peak.  Because this model does not distinguish between electron and positron beams, we expect this peak to be in the same location for both species.  This is indeed what we observe in the measured data.

\begin{figure}
   \centering
   \includegraphics*[width=0.9\textwidth]{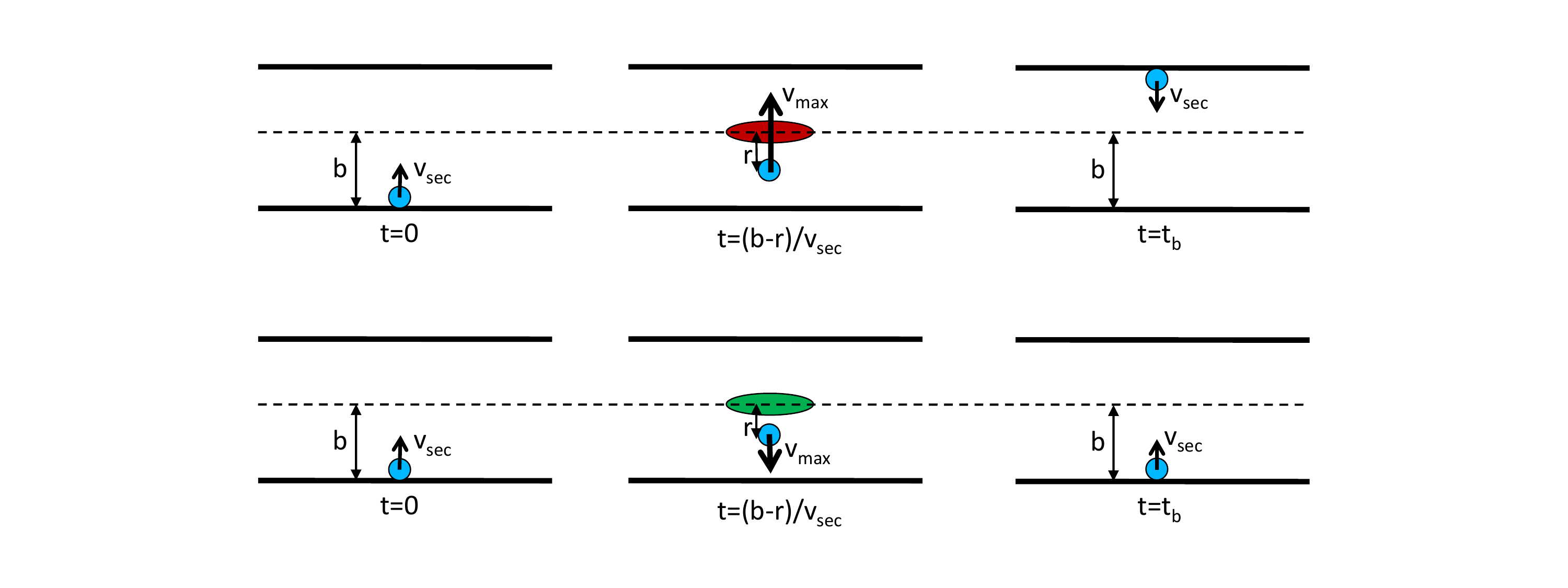}  
\caption{\label{fig:tb_diagram} Diagram of single bunch multipacting resonances: positron beams (top) and electron beams (bottom).  A secondary electron is released from the bottom wall (left), travels upward at speed $v_{sec}$, receives a kick from a passing bunch (middle), and hits the wall, releasing another secondary electron at time $t = t_b$ (right).}
\end{figure}

For the CESR dipole RFA ($b = 2.5$~cm),  the resonance should occur at 34~ns, which does not agree with either the electron or positron data.  In order to derive a more accurate prediction, we need to take into account the time it takes for a kicked electron to reach the chamber wall.  We define the resonant condition as the bunch spacing that results in an electron energy $E_2 = E_{max}$, where $E_{max}$ is the energy corresponding to peak secondary production (in eV).  This process is diagrammed in Fig.~\ref{fig:tb_diagram}.


The resonant condition now becomes:

\begin{align}
\begin{split}
t_b & = \frac{b - r}{v_{sec}} + \frac{b \pm r}{v_{max}} \\
v_{max} & \equiv \sqrt{\frac{2 q_e E_{max}}{m_e}} = \frac{2 c N_b r_e}{r} \pm v_{sec}
\label{eq:tb1_res2}
\end{split}
\end{align}

Here $r$ is the distance from the electron to the beam during the bunch passage, $N_b$ is the bunch population and $r_e$ is the classical electron radius.  Where there is a $\pm$ symbol, the plus sign applies for positron beams, and the minus for electron beams.



Eliminating $r$ from Eq.~(\ref{eq:tb1_res2}) and defining $k \equiv 2 c N_b r_e$ gives us a resonant bunch spacing (Eq.~(\ref{eq:tb1_res3})).  Interestingly, the condition is still the same for electron and positron beams.

\begin{equation}
t_{b,1} = \frac{b (v_{max} + v_{sec}) - k}{v_{max} v_{sec}}
\label{eq:tb1_res3}
\end{equation}


In this analysis we have used the impulse approximation for determining the beam kick~\cite{CERN:LHC97,doi:10.1142/S0217751X14300233}, which assumes that $r$ is much greater than the beam size. This approximation is valid as long as the distance from the electron to the beam is greater than a critical radius $r_c \approx 2 \sqrt{N_b r_e \sigma_z \sqrt{2/\pi}}$, where $\sigma_z$ is the bunch length.  For the conditions presented here, $\sigma_z \approx$ 17~mm, so the critical radius is 1.6~mm at 1~mA, and 2.9~mm at 3.4~mA.  For the resonant condition in Eq.~(\ref{eq:tb1_res2}), $r \approx$ 2.8~mm at 1~mA, and 9.6~mm at 3.4~mA.  So the impulse approximation is always valid, although it's close at low current.


\begin{figure}
   \centering
   \includegraphics*[width=0.9\textwidth]{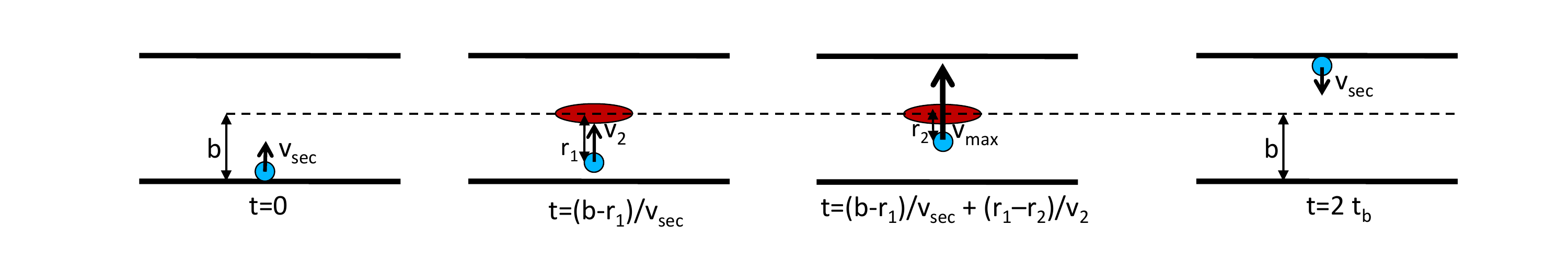}  
\caption{\label{fig:tb2_diagram} Diagram of a two-bunch multipacting resonance.  From left to right: a secondary electron is released from the bottom wall with speed $v_{sec}$.  It receives a kick from a passing bunch, and continues with higher velocity ($v_2$).  It is kicked again by a second bunch, bringing its speed up to $v_{max}$.  Finally, it hits the wall, releasing another secondary electron at time $t = 2 t_b$.}
\end{figure}

The 14~ns peak in the positron data is due to a higher order multipacting resonance, where it takes two bunches to set up the resonance condition. Here we consider the case where the first bunch gives some additional energy to the electron, so that it arrives near the center of the chamber in time for the second bunch, when it receives a large enough kick to give it energy $E_{max}$.  This process is shown in Fig.~\ref{fig:tb2_diagram}.

From this picture we can derive a system of equations for $t_{b,2}$ (where the subscript 2 is used to signify a 2-bunch resonance):

\begin{align}
\begin{split}
2 t_{b,2} &= \frac{b - r_1}{v_{sec}} + \frac{r_1 - r_2}{v_2} + \frac{r_2 + b}{v_{max}} \\
t_{b,2} &= \frac{r_1 - r_2}{v_2}  \\
v_2 &= v_{sec} + \frac{k}{r_1} \\
v_{max} &= v_2 + \frac{k}{r_2}
\label{eq:tb2_res}
\end{split}
\end{align}

Here $r_1$ is the distance between the beam and the electron during the first bunch passage, $r_2$ is this distance during the second bunch passage, and $v_2$ is the electron velocity after the first beam kick .  Note that this condition only applies to positron beams, since the kicks must be towards the beam.  These equations are a bit too unwieldily to be solved analytically, but they can be solved numerically to give predictions for the resonant bunch spacings.


\subsubsection{Comparison with Measured Data}

\begin{figure}
   \centering
   \begin{tabular}{c}
   \includegraphics*[width=0.6\textwidth]{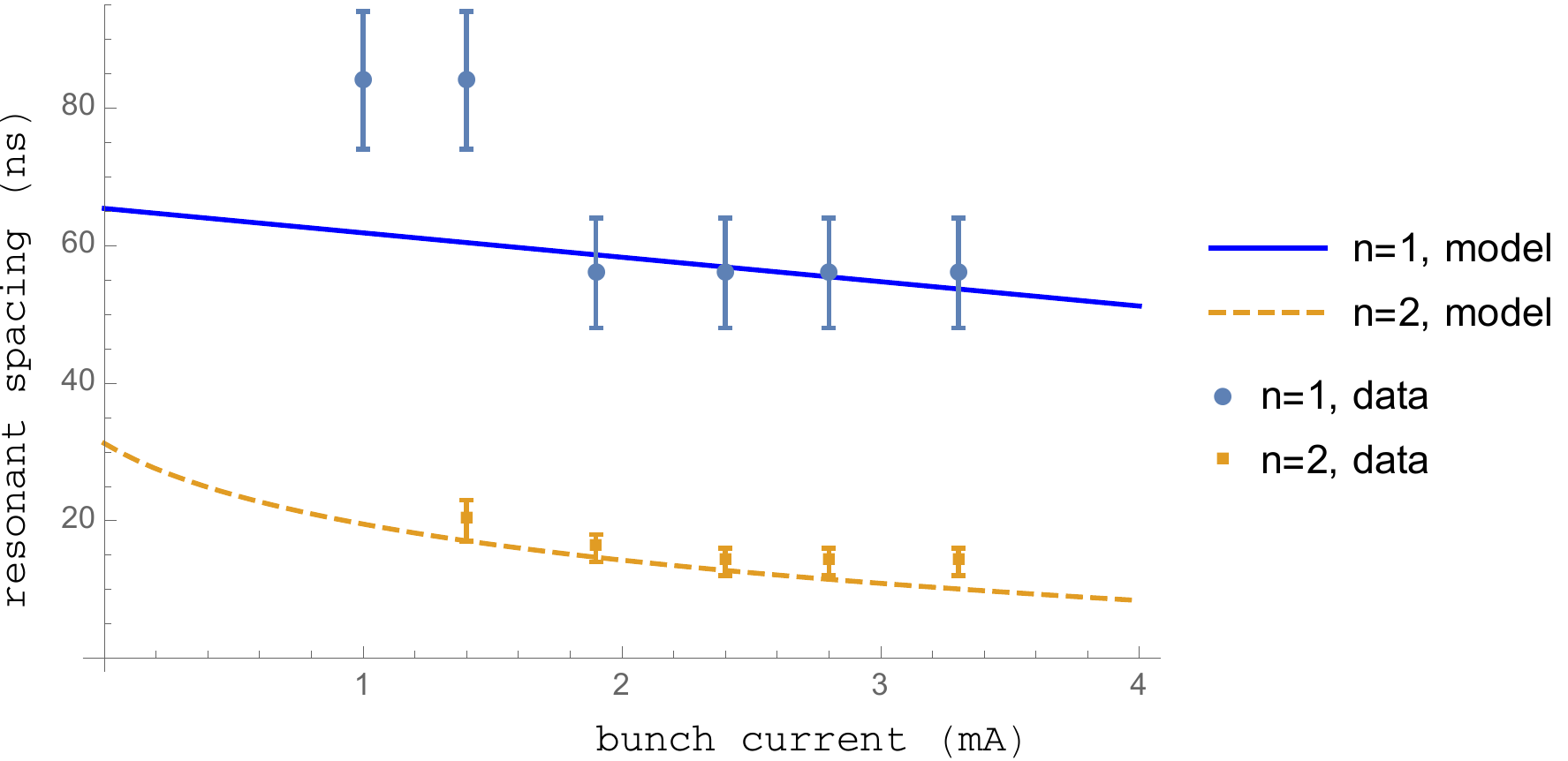} \\ 
   \includegraphics*[width=0.6\textwidth]{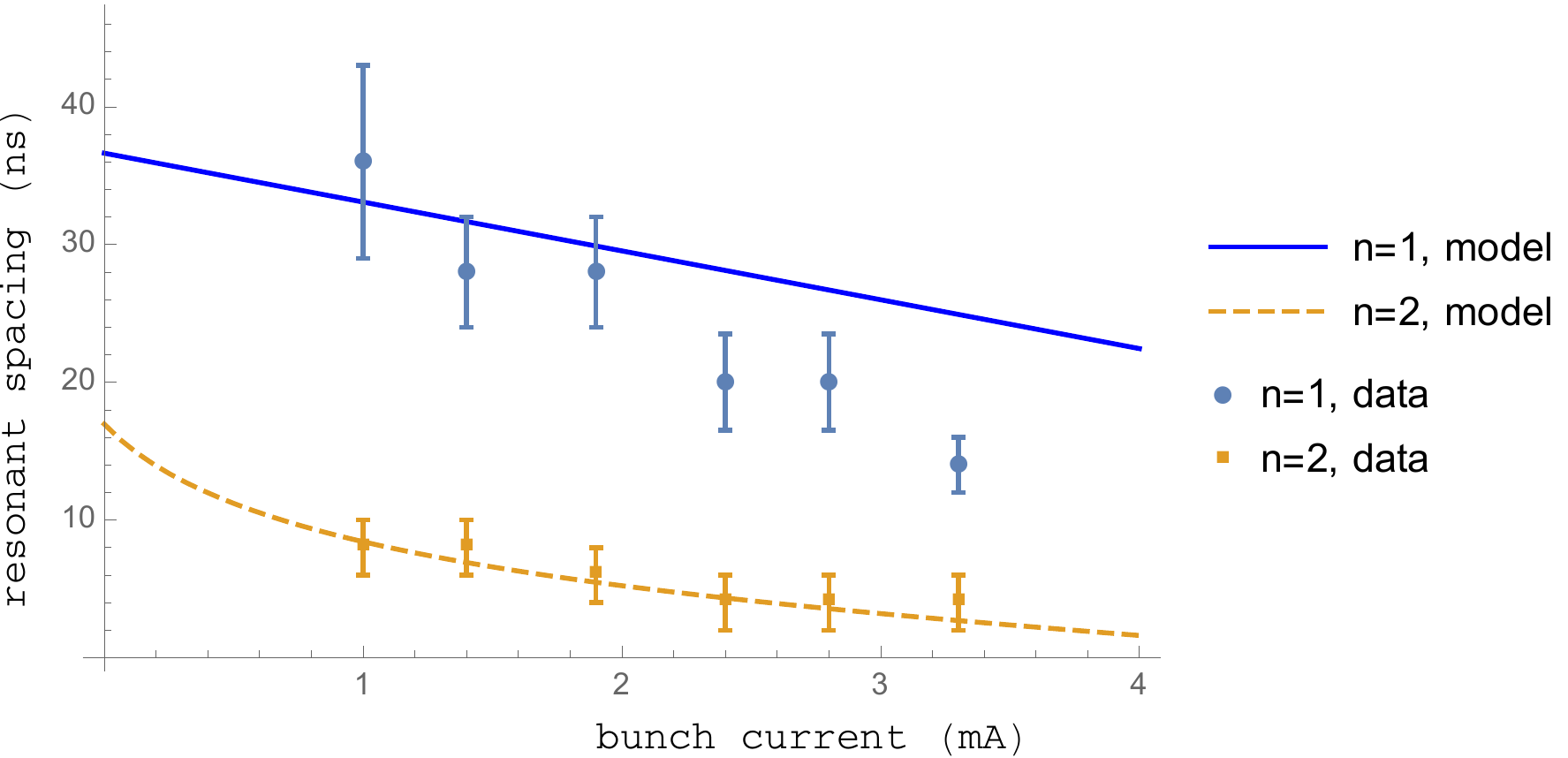} 
   \end{tabular}
\caption{\label{fig:mp_all} Comparison of measured and predicted multipacting resonances for the chicane (top) and CESR dipole (bottom) RFAs.  The solid lines represent 1-bunch resonances (Eq.~(\ref{eq:tb1_res3})), the dashed lines 2-bunch resonances (Eq.~(\ref{eq:tb2_res})), and the points are measured data.  In both cases the n=1 points are taken from the electron beam data, and the n=2 from the positron data.  The error bars are defined as half the difference in bunch spacing between successive measurements.}
\end{figure}

Fig.~\ref{fig:mp_all} compares the measured and predicted resonances for both the chicane and CESR dipole chambers.  Effectively, we have varied the two most important parameters of the model: bunch current, and chamber size (since the two dipole RFAs have different chamber heights).    Overall there is good agreement between the data and model for all measured resonances.  In particular, the model captures the major features of the data:

\begin{itemize}
    \item For the chicane RFA, the 1-bunch resonance appears in both the electron and positron data, at the same bunch spacing.
    \item The 2-bunch resonances are only observed in the positron data, at lower spacing than the 1-bunch resonances. 
    \item All resonances move toward lower bunch spacing at higher current.
\end{itemize}

The 1-bunch resonance is not seen as clearly in the CESR dipole RFA positron data, though a ``shelf" can be seen at 1.4~mA, which does correspond to the electron data peak.  Simulations (Section~\ref{ssec:mult_sim}) also predict a peak.  The lack of a clear resonance in the data may be a result of the depletion phenomena described in a previous paper (\cite{NIMA770:141to154}, Sec. 3.1.2).  Essentially, in a strong field (such as the 2~kilogauss field of the CESR dipole RFA), the RFA can actually become less sensitive to multipacting, since it depletes the cloud under the RFA holes, exactly where it's measuring.  In general the 1-bunch resonances are less pronounced than the 2-bunch resonances; this may be why we still see the 2-bunch resonance.



 The model and data are also in quantitative agreement, with two exceptions: the 1-bunch resonance for the chicane dipole at low current, and the 1-bunch resonance for the CESR dipole at high current.  The former discrepancy may be due to the impulse approximation not being valid (as explained above).  The latter discrepancy may be due to the fact that we are ignoring the beam's image charge, and the cloud's space charge.  The chicane RFA chamber is in a circular chamber, so there will be no image charge (assuming a centrally located beam).  It is also located in a long straight section that receives relatively little synchrotron radiation.  This means the overall cloud density is lower, and space charge is less important.  The CESR dipole chamber, however, is (approximately) elliptical, so image charge can be important.  It is also located in a high radiation environment.  An improved model, which takes image charge and space charge into account, would probably fit this data better.



Measurements of multipacting resonances with a positron beam at the Advanced Photon Source~\cite{PRSTAB6:034402} found a peak at 20~ns for bunch populations in the range of $3.45\times10^{10}$ to $5.75\times10^{10}$.  Plugging these numbers and the chamber half-height (21~mm) into Eq.~(\ref{eq:tb1_res3}) gives a resonant spacing of 18-23~ns, consistent with their result.  However, they measured a different resonance (30~ns) for an electron beam, which is not predicted by our theory.  Their measurements were made in a field-free region, with an RFA located at an angle with respect to the top of the chamber, so our one-dimensional model may not be completely valid.  Nonetheless it is suggestive that the location of the positron peak agrees with our prediction.

Table~\ref{tab:res_other} lists the predicted locations of multipacting resonances for some proposed accelerators with positively charged beams.  Also included for comparison are the two most common operating modes of the APS (which now uses electron beams, so there is no 2-bunch resonance).  The LHC is not included, because the beam is so intense that $E_{2} > E_{max}$ at the beam pipe wall, so the machine will generate high energy secondaries regardless of bunch spacing.


\begin{table}
  \centering
  \setlength\tabcolsep{5mm}
  \caption{Resonant bunch spacings ($t_{b,1}$, $t_{b,2}$) compared to operational spacing ($t_b$) for different accelerators.}\label{tab:res_other}
  \begin{tabular}{c c c c c c}
     \hline \hline
     Machine    & $N_b$              & b (cm) & $t_b$ (ns) & $t_{b,1}$ (ns) & $t_{b,2}$ (ns) \\
     \hline
     ILC DR     & $2\times10^{10}$   & 2.5    & 6        & 32        & 7.6   \\
     CLIC DR    & $4.1\times10^{9}$  & 3      & 0.5      & 43        & 17.4 \\
     SuperKEKB       & $9\times10^{10}$   & 4.5      & 4        & 46        & 5.4   \\
     APS (324b) & $7.1\times10^{9}$  & 2.1    & 11       & 29        & X          \\
     APS (24b)  & $9.5\times10^{10}$ & 2.1    & 153      & 9         & X          \\
     \hline \hline
   \end{tabular}
\end{table}

It is worth noting that running with very short bunch spacing (as many cutting edge accelerators do) can actually be advantageous from an electron cloud point of view, since it avoids both multipacting resonances.  Running with high current and very large bunch spacing (as some light sources do) also works.  However, it is important to keep in mind that this model does not include the cloud's space charge, which could be an important effect in these high intensity machines.  Particle tracking simulations (see Section~\ref{ssec:mult_sim}) can be used to more accurately predict the resonances.


\subsection{\label{ssec:cyc_res} Cyclotron Resonances}

\begin{figure}
   \centering
   \begin{tabular}{c}
   \includegraphics*[width=0.6\textwidth]{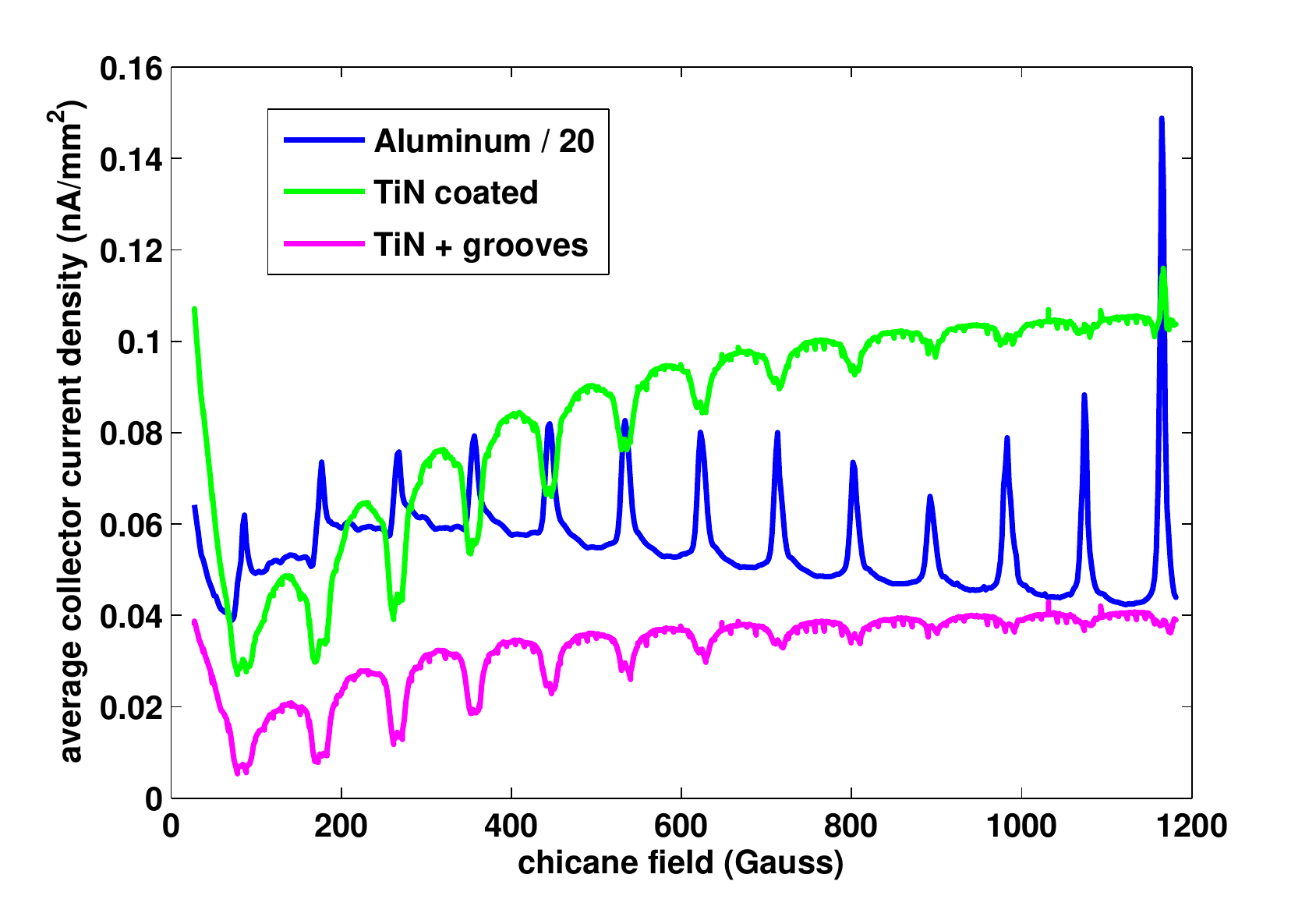} \\ 
   \includegraphics*[width=0.6\textwidth]{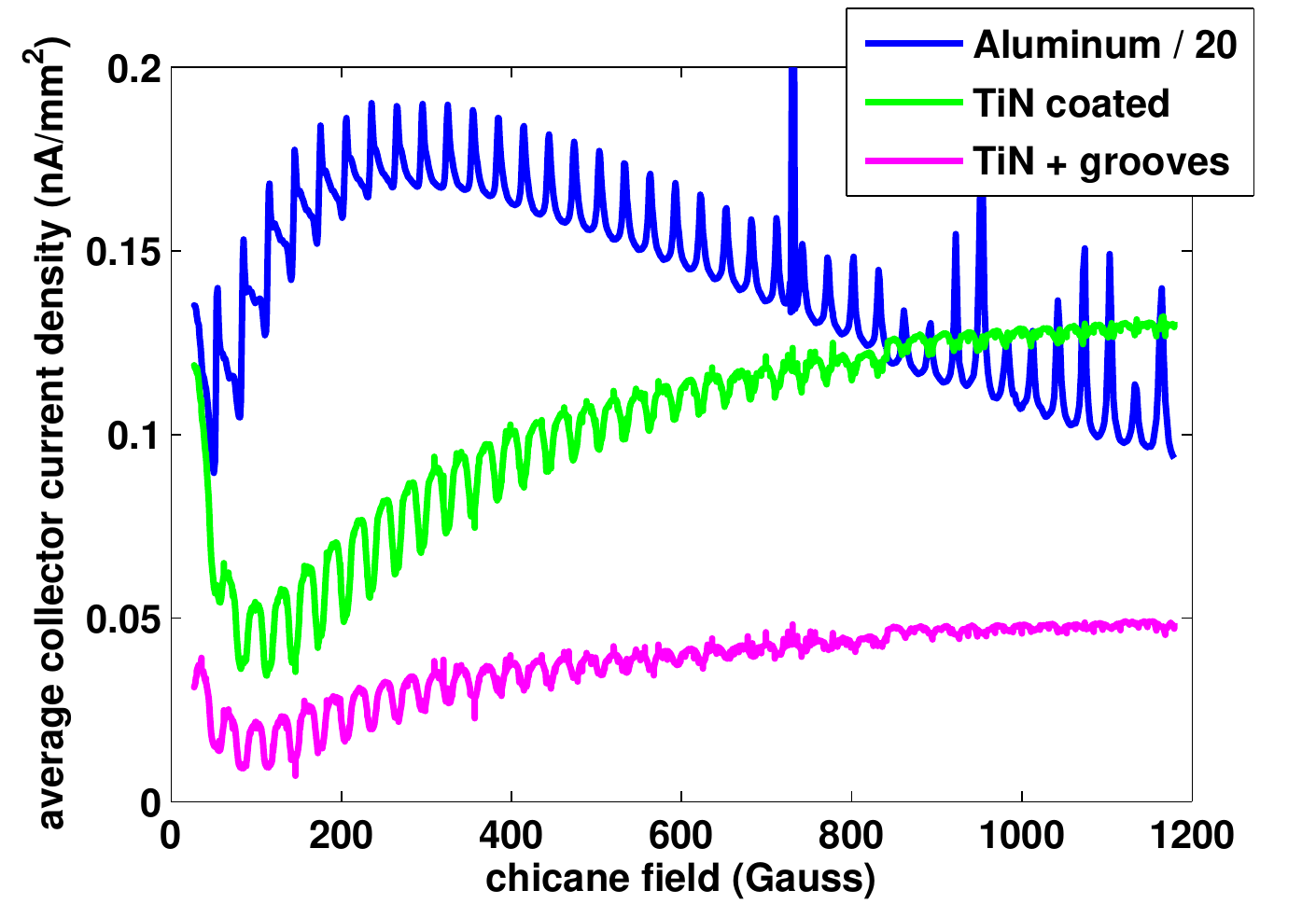} \\ 
   \end{tabular}
   \caption{\label{fig:chicane_scan} RFA signal as a function of chicane magnetic field: 1x45x1~mA $e^+$, 5~GeV.  Top: 4~ns spacing.  Bottom: 12~ns spacing.  Cyclotron resonances are observed every 89~gauss with 4~ns spacing, and every 30~gauss with 12~ns spacing, as predicted by Equation~(\ref{eq:cyc}).  Note that the aluminum chamber signal is divided by 20.}
\end{figure}

By varying the strength of the chicane magnets, we can also study the behavior of the cloud at different dipole magnetic field values.  Fig.~\ref{fig:chicane_scan} shows RFA data taken as a function of magnetic field strength, at two different bunch spacings.  The most prominent feature of the data is regularly occurring spikes or dips, which are seen in all cases.  These correspond to ``cyclotron resonances," which occur whenever the cyclotron period of cloud electrons is an integral multiple of the bunch spacing (see Section~\ref{ssec:cyc}).  For 4~ns bunch spacing we expect them every 89~gauss; and for 12~ns spacing, every 30~gauss.  This is exactly what is seen in the data.  Another interesting feature of this measurement is that these resonances appear as peaks in the RFA signal in the aluminum chamber, but as dips in the coated chambers.  This difference in the behavior of the two chamber materials is explained in Section~\ref{ssec:cyc_res_sim}.



\subsection{\label{ssec:tramp} Anomalous Enhancement}

\begin{figure}
   \centering
   \includegraphics*[width=0.6\textwidth]{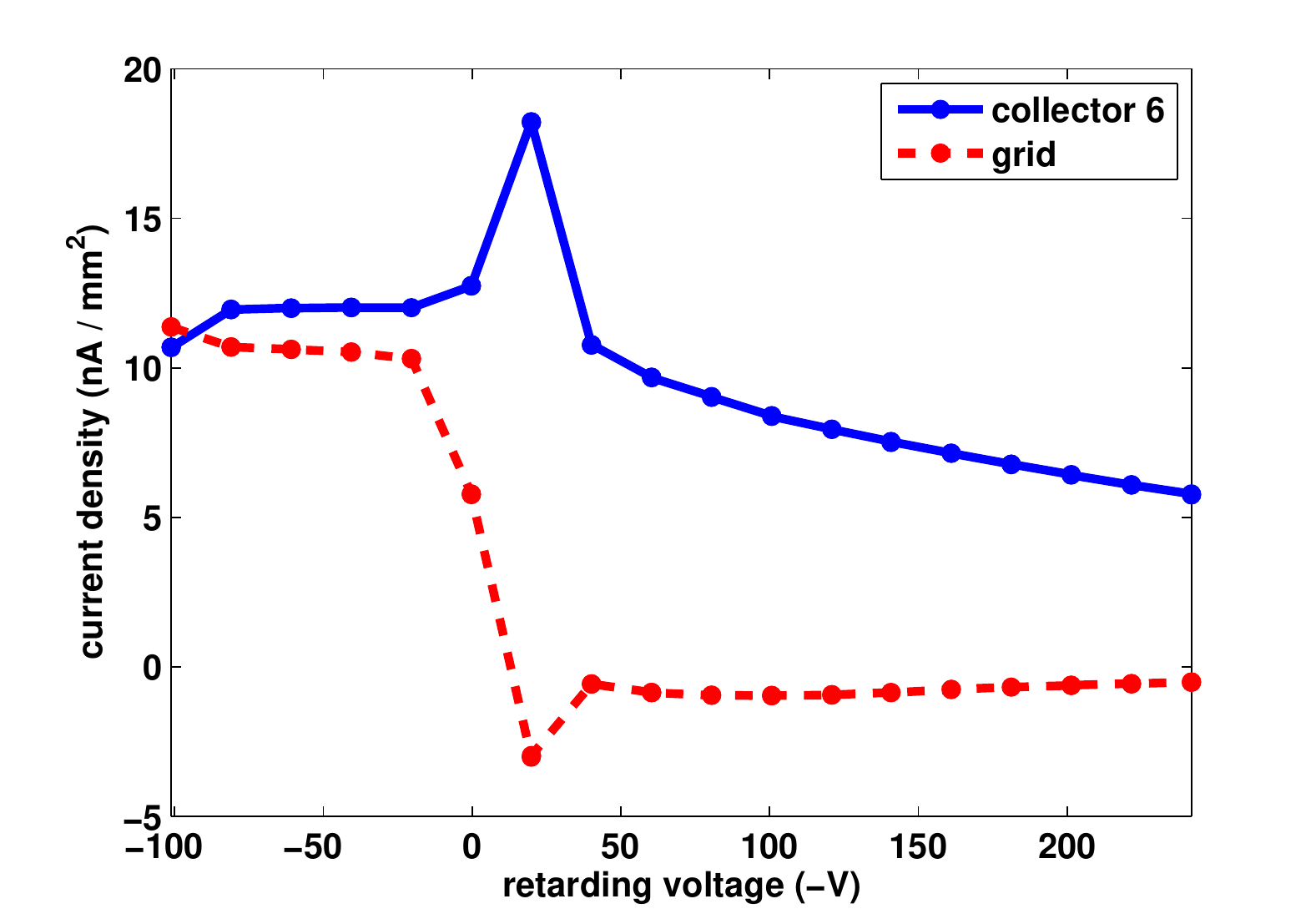}
   \caption[Resonant enhancement in wiggler data]{\label{fig:tramp_example} Resonant enhancement in wiggler data, 45 bunches, 1.25 mA/bunch, $e^+$, 2.1~GeV, 14~ns.  Note that there are 12 collectors, so collector 6 is one of the central ones.}
\end{figure}

Detailed analysis of the wiggler RFA data is complicated by an interaction between the cloud and the RFA itself.  Fig.~\ref{fig:tramp_example} shows a voltage scan done with an RFA in the center pole of a wiggler (approximated by a 1.9 T dipole field).  Here one can see a clear enhancement in the signal at low (but nonzero) retarding voltage.  Since the RFA should simply be collecting all electrons with an energy more than the magnitude of the retarding voltage, the signal should be a monotonically decreasing function of the voltage.  So the RFA is not behaving simply as a passive monitor.   A similar effect has been observed in a strong dipole field at KEKB~\cite{NIMA598:372to378}.  The spike in collector current is accompanied by a corresponding dip in the grid current, suggesting that the grid is the source of the extra collector current.


This spurious signal comes from a resonance between the bunch spacing and retarding voltage.  To understand this, consider an electron which collides with the retarding grid and generates a secondary.  Because electrons are so strongly pinned to the magnetic field lines in a 1.9~T field, this electron is likely to escape through the same beam pipe hole through which it entered. An electron ejected from the grid will gain energy from the retarding field before it re-enters the vacuum chamber.  If it is given the right amount of energy, it will be near the center of the vacuum chamber during the next bunch passage, and get a large beam kick, putting it in a position to generate even more secondaries.  This process, which we have dubbed the ``trampoline effect", is essentially an artificial multipacting resonance.  If we take Eq.~(\ref{eq:tb1_res}) from Section~\ref{ssec:mult_res}, and use the retarding voltage in place of the secondary electron energy, the resonance conditions becomes:

\begin{equation}
\label{eq:tramp}
V_{ret} = \frac{m_e b^2}{2 q_e t_b^2}
\end{equation}

\begin{figure}
   \centering
   \includegraphics*[width=0.6\textwidth]{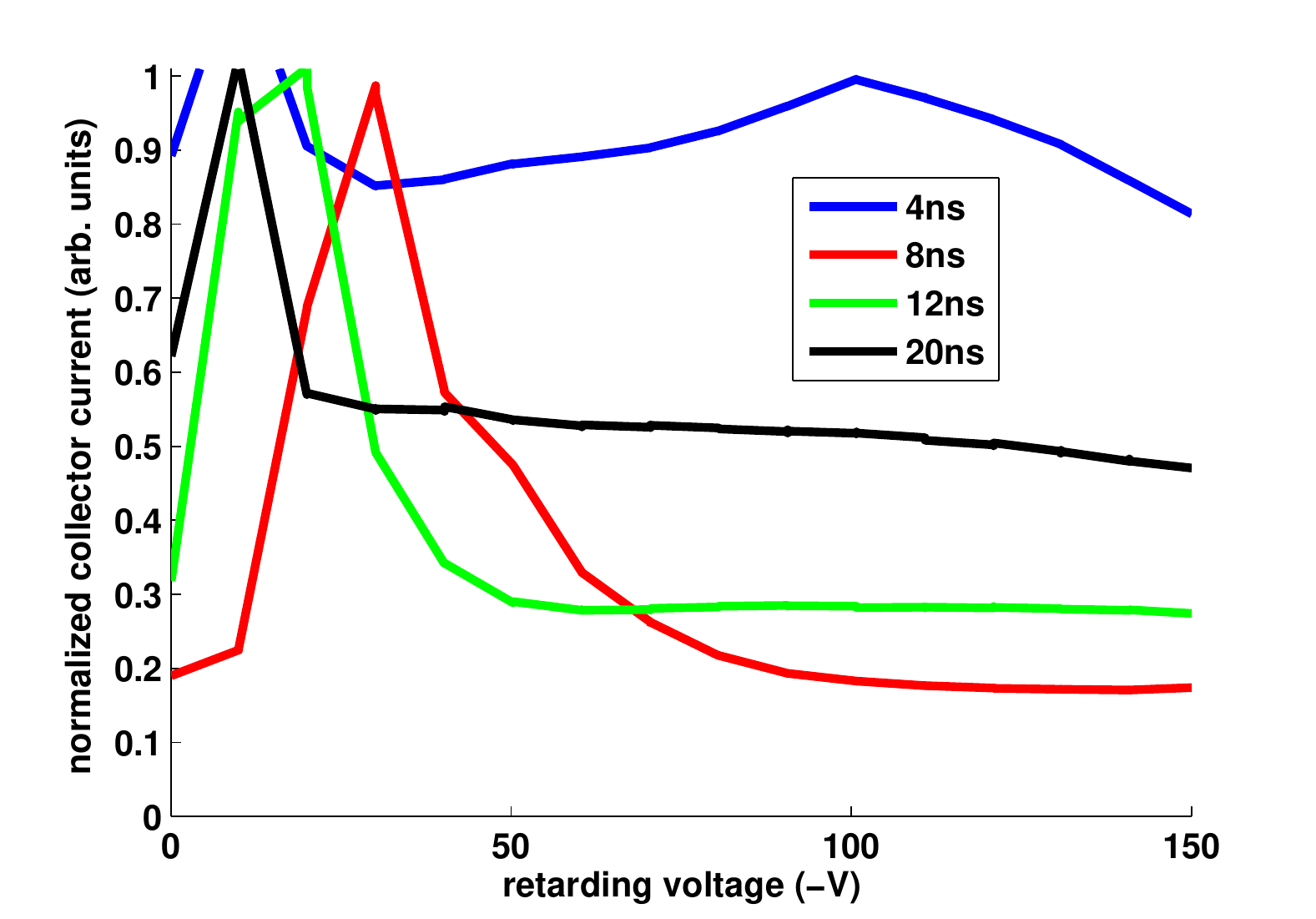}
   \caption[Resonant spike location at different bunch spacings]{\label{fig:tramp_spacing} Resonant spike location at different bunch spacings, 1x45x1.25 mA $e^+$, 5~GeV.  Only the signal in the central collector is plotted.}
\end{figure}

Here $V_{ret}$ is the retarding voltage, $b$ is the chamber half-height, $t_b$ is the bunch spacing, $m_e$ is the electron mass, and $q_e$ is the electron charge.  Fig.~\ref{fig:tramp_spacing} plots a series of retarding voltage scans done with a wiggler RFA, for 4, 8, 12, and 20~ns bunch spacing.  The trampoline effect is seen in all cases, with the spike occurring at $\sim$110, 30, 15, and 10~V, respectively.  Meanwhile, the simple model given in Eq.~(\ref{eq:tramp}) predicts 111, 28, 12, and 4~V, respectively.  The predictions are quite close to the measurements, especially for short bunch spacing.  The second spike at low voltage in the 4~ns data corresponds to a two-bunch resonance, also described in Section~\ref{ssec:mult_res}.

\section{Simulations}

While the analytical models described above are generally successful at explaining our data, additional insight can be gained by using more detailed computer simulations.  The results presented here were obtained with the particle tracking code POSINST~\cite{MBI97:170,LHC:ProjRep:180,PRSTAB5:124404}.  In POSINST, a simulated photoelectron is generated on the chamber surface and tracked under the action of the beam.  Secondary electrons are generated via a probabilistic process.  Space charge and image charge are also included in the simulation.

\subsection{\label{sec:modeling} RFA Modeling}

In order to accurately predict the RFA signal, a sophisticated model of the detector must be incorporated into the code.  Our model has been described in detail for the RFAs installed in field free regions~\cite{PRSTAB17:061001}; the dipole RFA models are essentially the same.  In short, when a macroparticle in the simulation collides with the vacuum chamber wall in the region covered by the RFA, a special function is called which calculates a simulated RFA signal based on the particle's incident energy and angle.  The signal is binned by energy and transverse position, reproducing the energy and position resolution of the RFA.

\begin{figure}
\centering
\includegraphics[width=.6\linewidth]{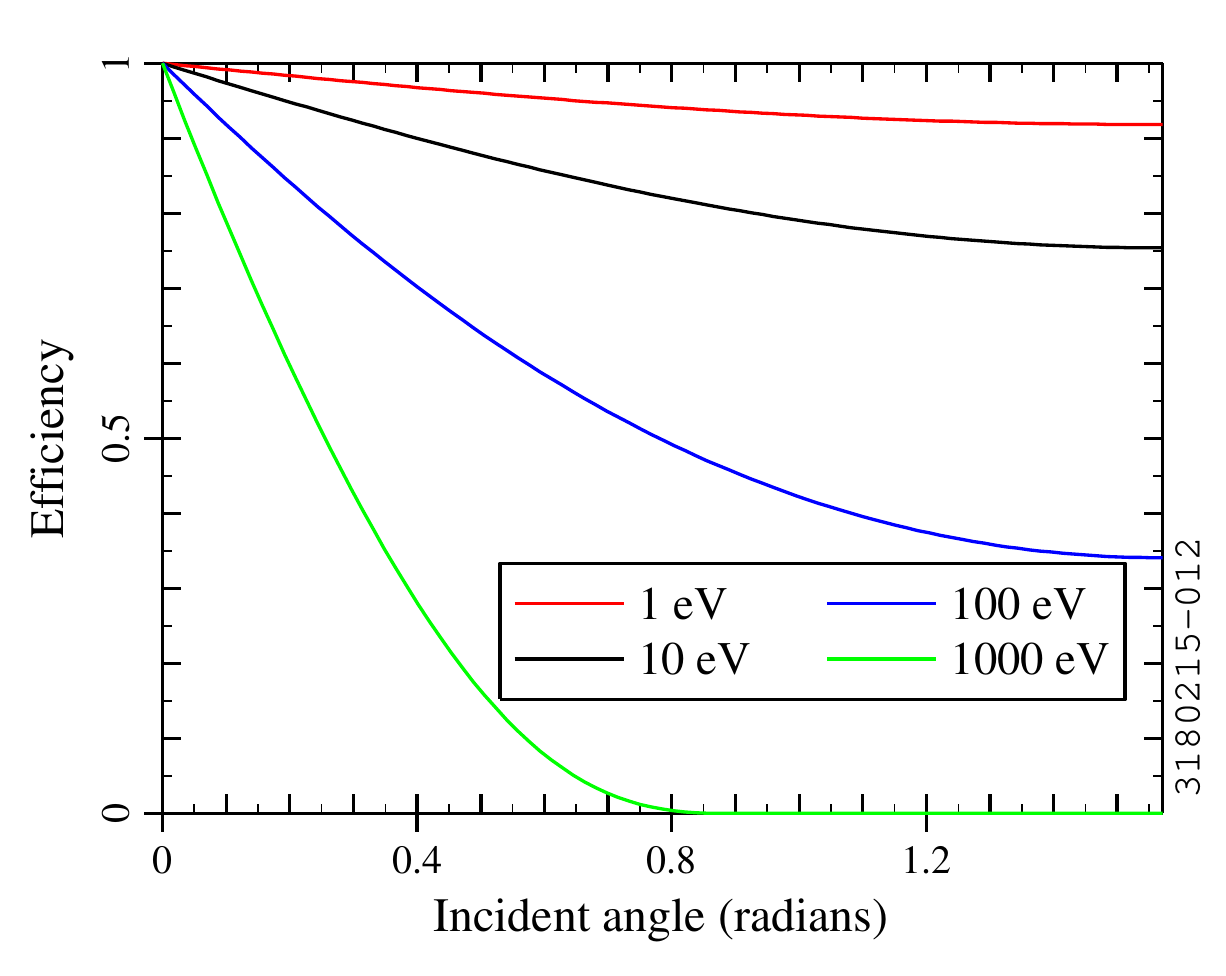} \\ 
\caption[Simulated RFA efficiency vs. incident angle, dipole RFA]{\label{fig:dipole_rfa_eff} Simulated RFA efficiency vs. incident angle for the chicane dipole RFA, with a 810~gauss magnetic field.}
\label{dipoleff}
\end{figure}

Fig.~\ref{fig:dipole_rfa_eff} shows the efficiency (fraction of the macroparticle's charge that contributes to the RFA signal) as a function of incident angle in the chicane RFA.  This represents the probability that an incoming electron will make it through the beam pipe hole and grids, and to the collector.  Note that low energy particles have a very high efficiency, due to their small cyclotron radius.

\begin{figure}
\centering
\includegraphics[width=.6\linewidth]{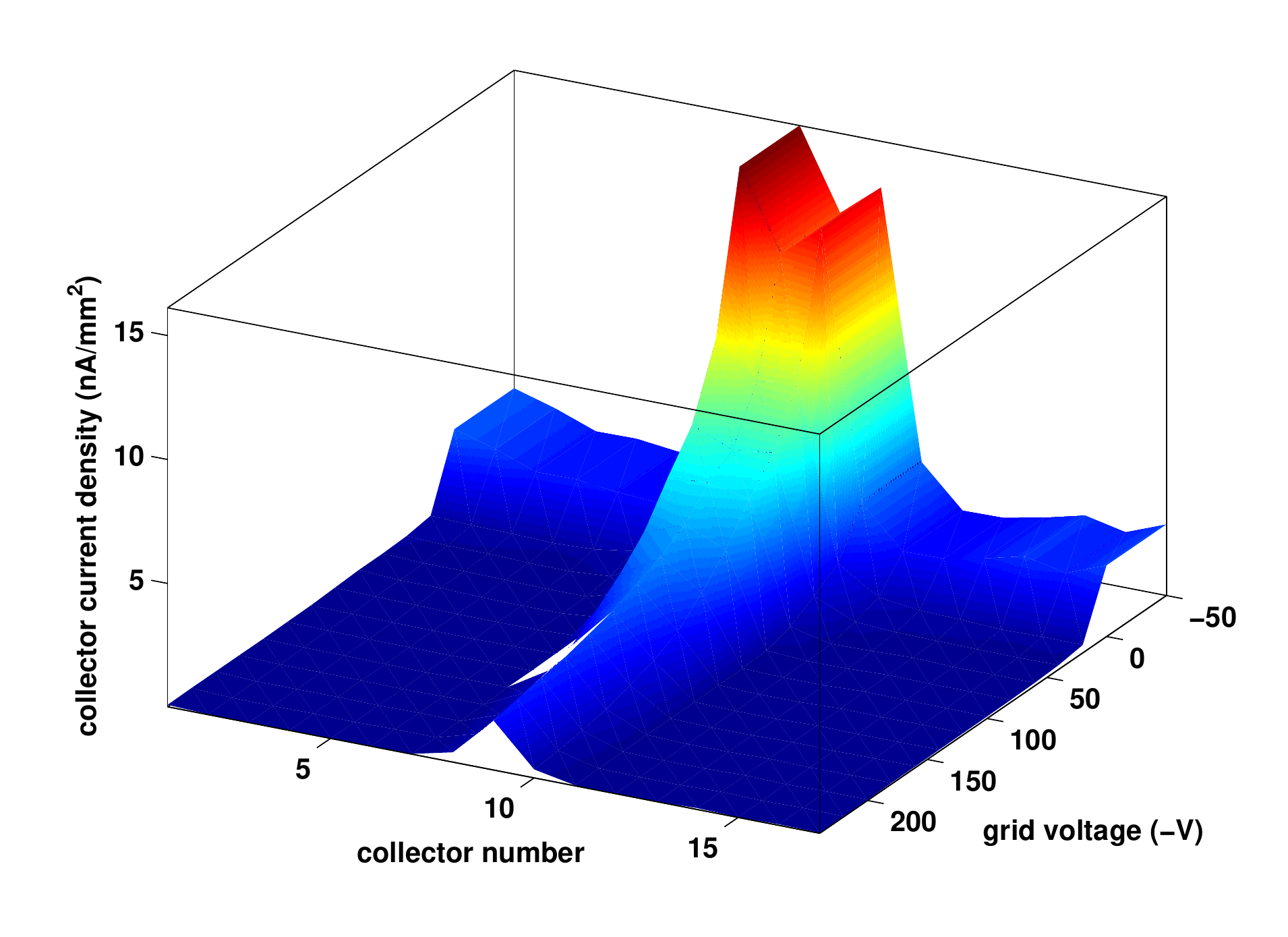}
\caption{\label{fig:sim_example} Example aluminum chicane RFA simulation: 1x45x1.25~mA $e^+$, 14~ns, 5.3~GeV.  Compare to Fig.~\ref{fig:chic_dipole_meas}.}
\end{figure}

Using the model described above, we ran simulations for the dipole RFAs, for various beam conditions.  Fig.~\ref{fig:sim_example} shows a typical example, for the aluminum chicane RFA.  Overall, the agreement with data (Fig.~\ref{fig:chic_dipole_meas}) is reasonable, without any additional tuning of the simulation parameters.

\subsection{\label{ssec:mult_sim} Simulation of Multipacting Resonances}

Because the simulation contains all the relevant features of our multipacting model (i.e. secondary emission, beam kicks, chamber geometry), it should be able to reproduce the resonances predicted by the model.  In addition, we are able to vary the secondary emission energy, to study the effect this has on the resonant spacings.  According to Eq.~(\ref{eq:tb1_res}), the 1-bunch resonance should have an approximately inverse dependence on the emission velocity, i.e. $t_{b,1} \sim 1/\sqrt{E_{sec}}$.  The 2-bunch resonance should have a much weaker dependence on emission energy.

\begin{figure*}
\centering
\begin{tabular}{c c}
\includegraphics[width=.5\linewidth]{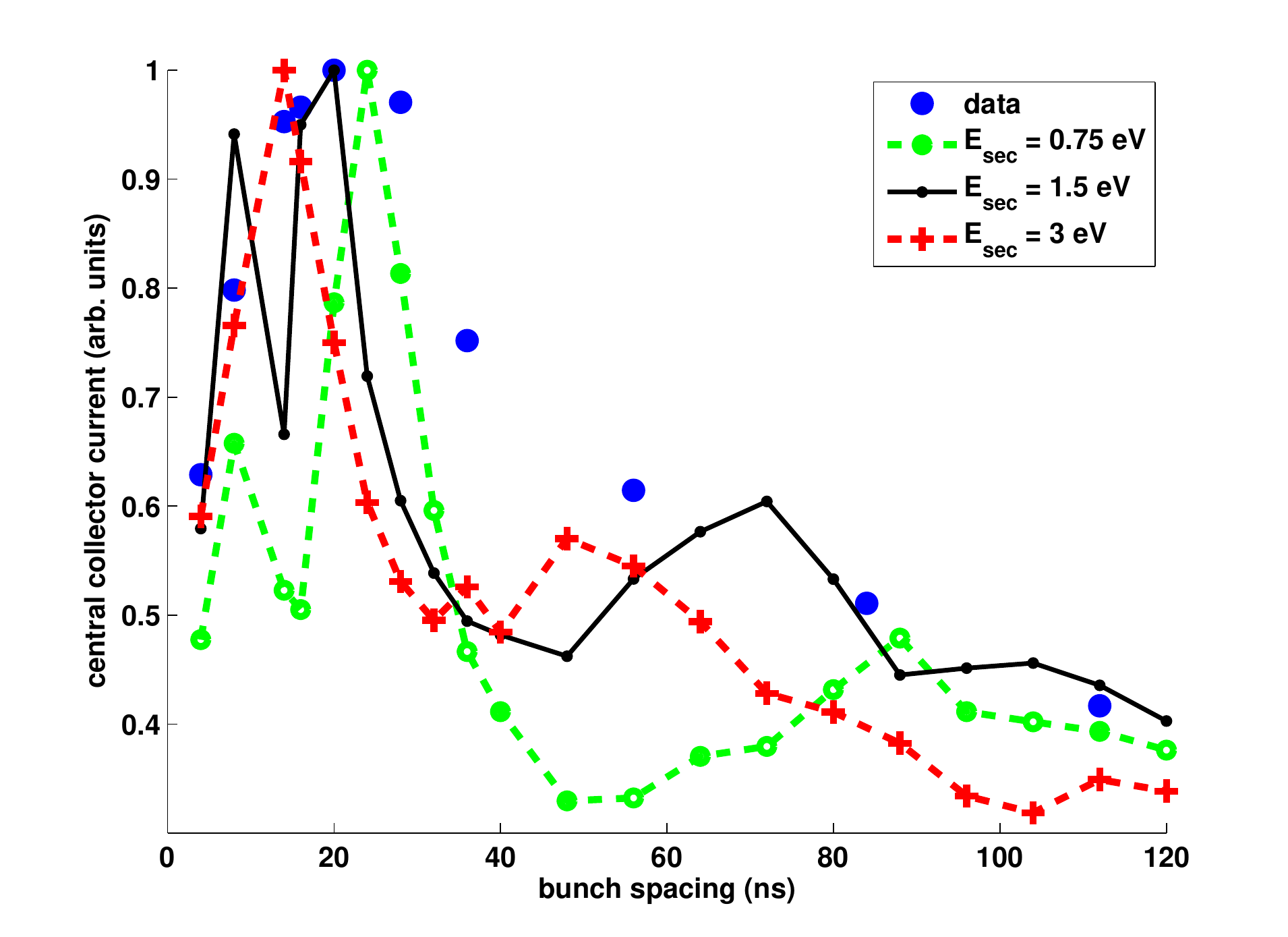} 
\includegraphics[width=.5\linewidth]{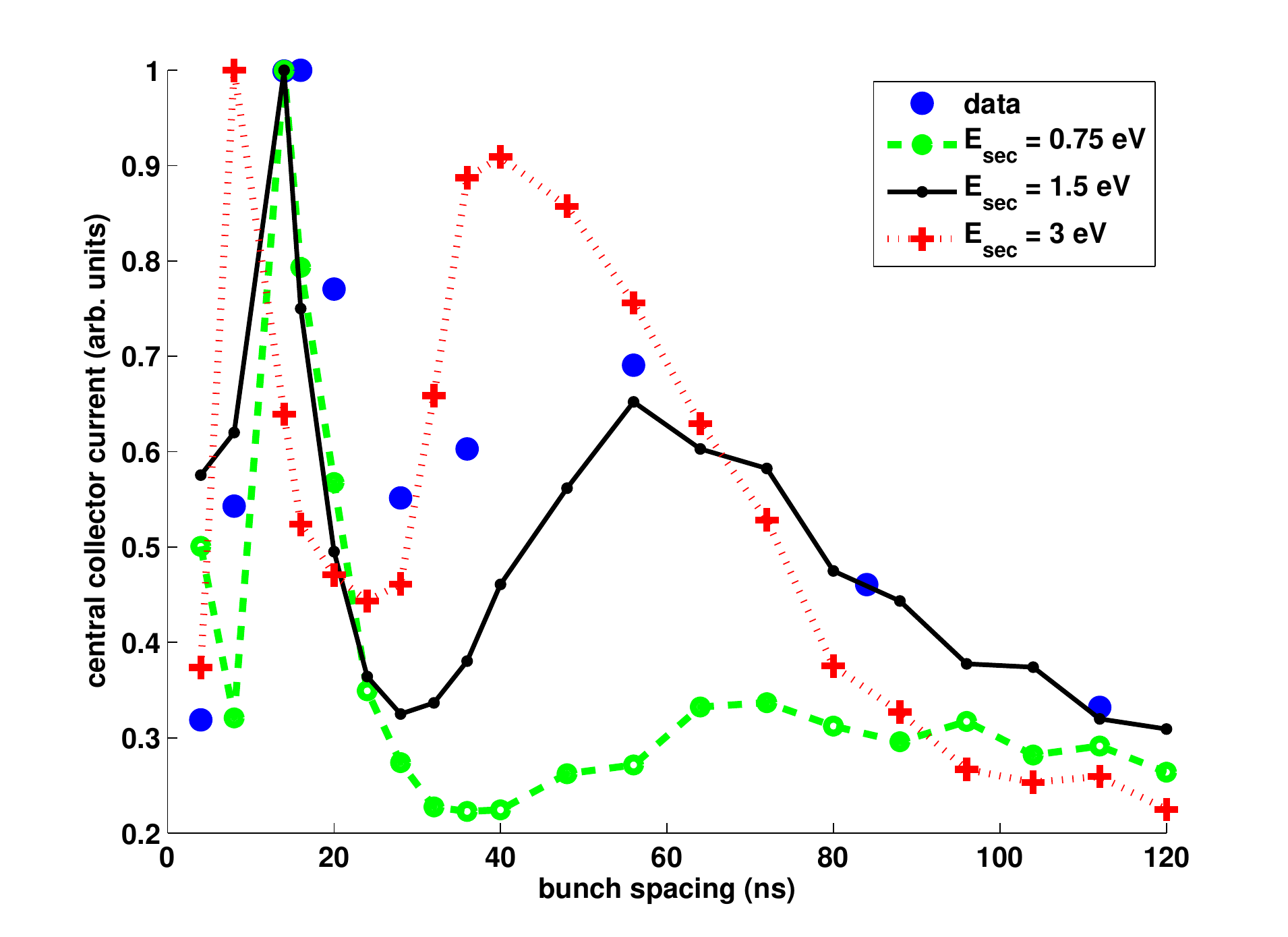} \\ 
\includegraphics[width=.5\linewidth]{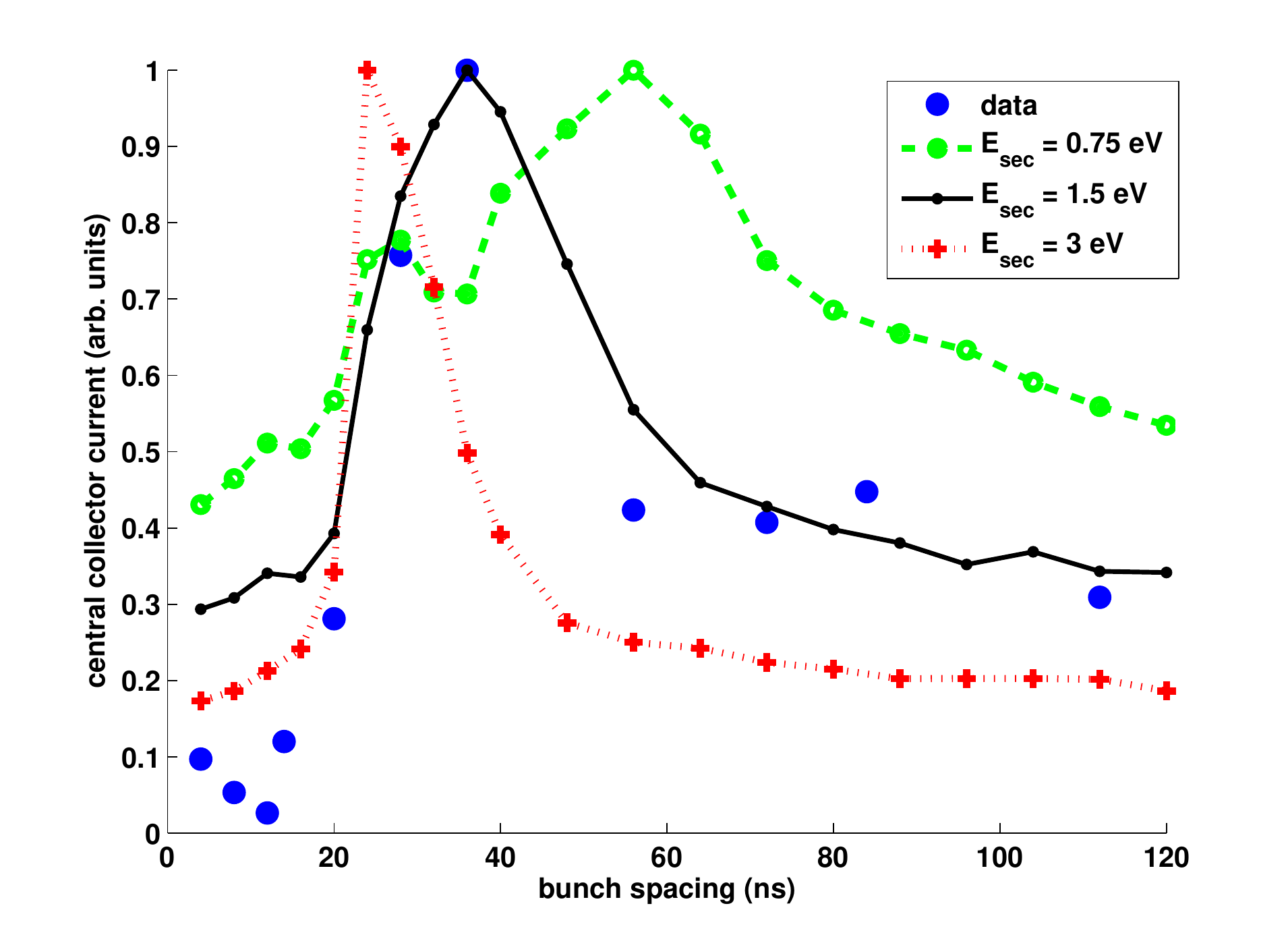} 
\includegraphics[width=.5\linewidth]{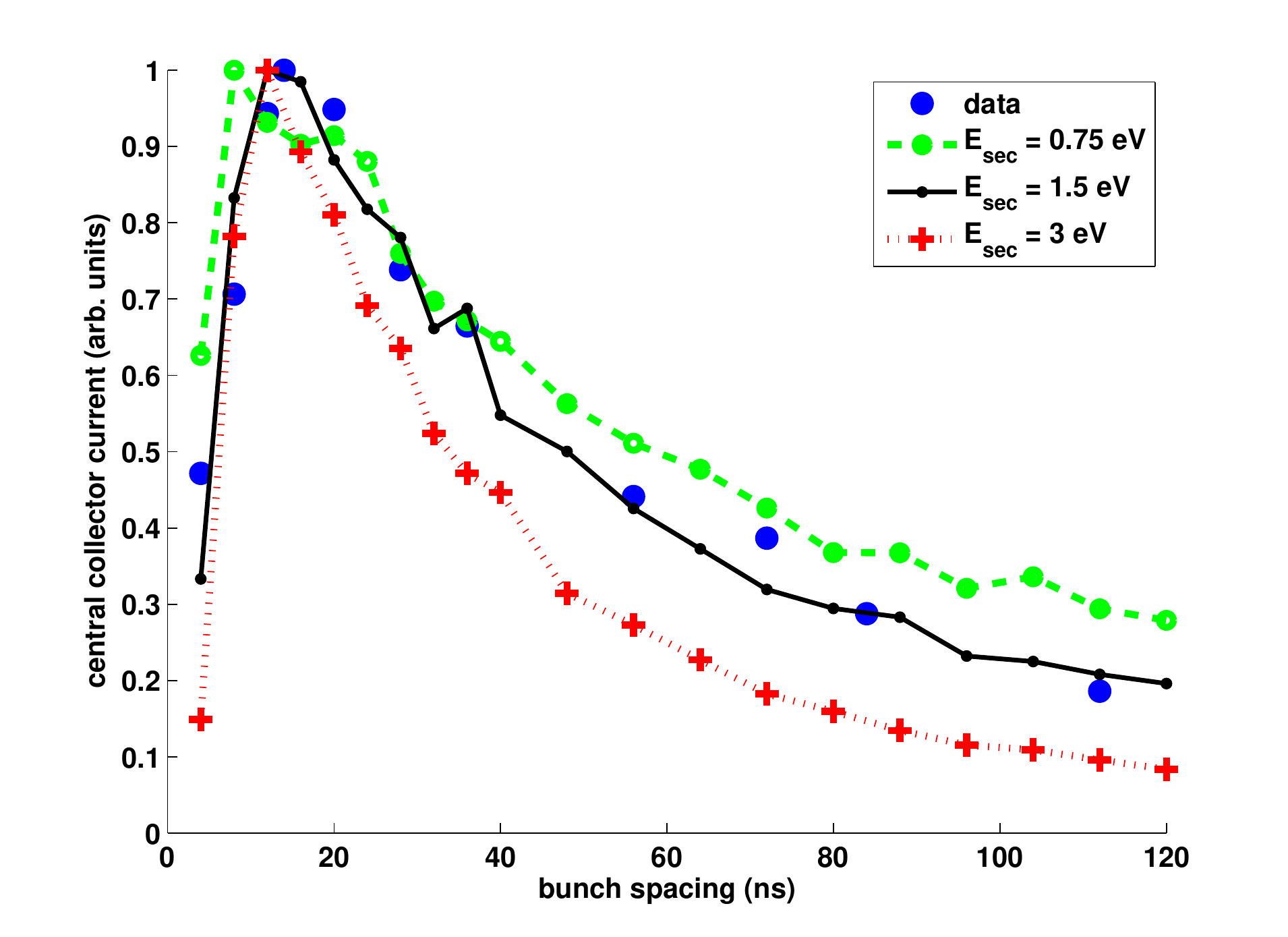} \\ 
\end{tabular}
\caption{\label{fig:dip_spacing_sim} Simulation of the multipacting resonances, compared to measurements, for different secondary emission energies. Top left: chicane RFA, 1.4~mA, e$^+$; top right: chicane RFA, 3.4~mA, e$^+$; bottom left: CESR dipole RFA, 1~mA e$^-$; bottom right: CESR dipole RFA, 3.4~mA, e$^-$.  All cases were done with 20 bunches, at beam energy 5.3~GeV.}
\end{figure*}

Fig.~\ref{fig:dip_spacing_sim} plots the simulated central collector signal as a function of bunch spacing, for four different combinations of chamber, bunch current, and beam species.  Both the 1-bunch and 2-bunch multipacting peaks are observed.  As predicted by the model, the locations of these peaks (especially for the 1-bunch resonance) are sensitive to the energy spectrum of emitted secondary electrons.  A secondary emission energy distribution peaked at 1.5~eV is generally consistent with the data, in particular with the locations of the multipacting peaks.  Lowering the emission energy to 0.75~eV moves the peaks to higher bunch spacings, and broadens the peaks.  Increasing the energy to 3~eV moves the peaks to lower spacings, and also results in narrower peaks.  Neither of these cases are consistent with the measured data.  Thus this comparison provides a fairly sensitive indirect measurement of the secondary emission energy.

In general, the data, analytical model, and simulation are in good agreement, assuming 1.5~eV secondary electrons.  It is notable that the simulation agrees well with the high current electron beam data in the CESR chamber (which the analytical model did not match well).  This is most likely because the simulation includes space and image charge, which are important in the high current regime.


For the sake of simplicity, the angular distribution of emitted secondaries was set to be strongly peaked at normal to the vacuum chamber wall (POSINST parameter \texttt{pangsec}~\cite{PRSTAB5:124404} was set to 10).  This was done to make it easy to compare the location of resonances to those predicted by the model.  In reality the electrons should be emitted at various angles, which would complicate the analysis, but may give a qualitatively better fit to the data.  Studying the effect of \texttt{pangsec} and other simulation parameters on these results would be an interesting subject for future study.






\subsection{\label{ssec:cyc_res_sim} Simulation of Cyclotron Resonances}

\begin{figure*}
\centering
\begin{tabular}{c c}
\includegraphics[width=.45\linewidth]{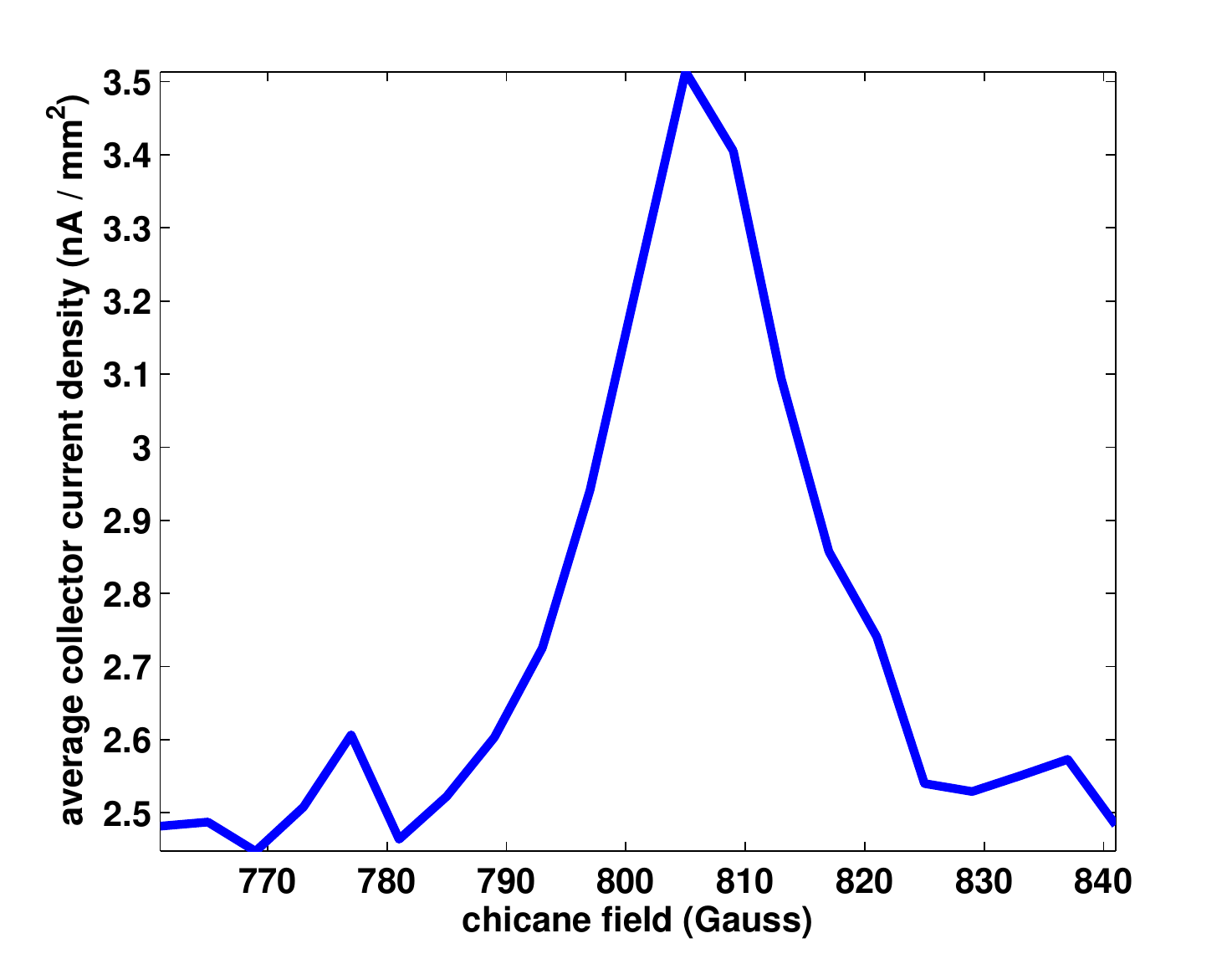}
\includegraphics[width=.45\linewidth]{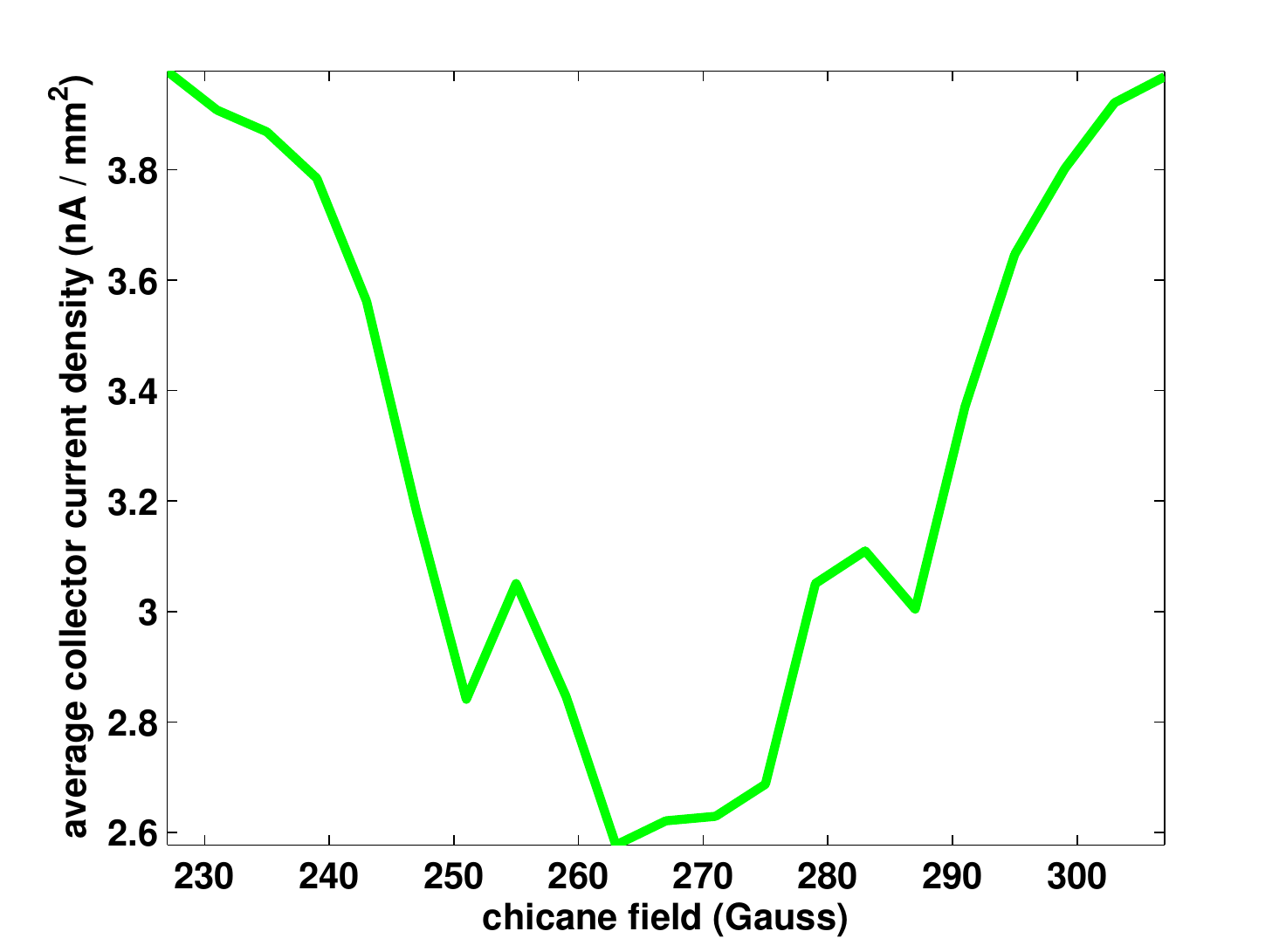} \\
\end{tabular}
\caption[Simulation of cyclotron resonances]{\label{fig:cyc_sim} Simulation of cyclotron resonances observed by an RFA in aluminum (left) and TiN (right) chambers, 1x45x1~mA $e^+$, 4~ns, 5~GeV. Note that, as in Fig.~\ref{fig:chicane_scan}, the resonance appears as an increase in the aluminum chamber signal, but a decrease in the TiN chamber signal.}
\end{figure*}

\begin{figure}
\centering
\includegraphics[width=.6\linewidth]{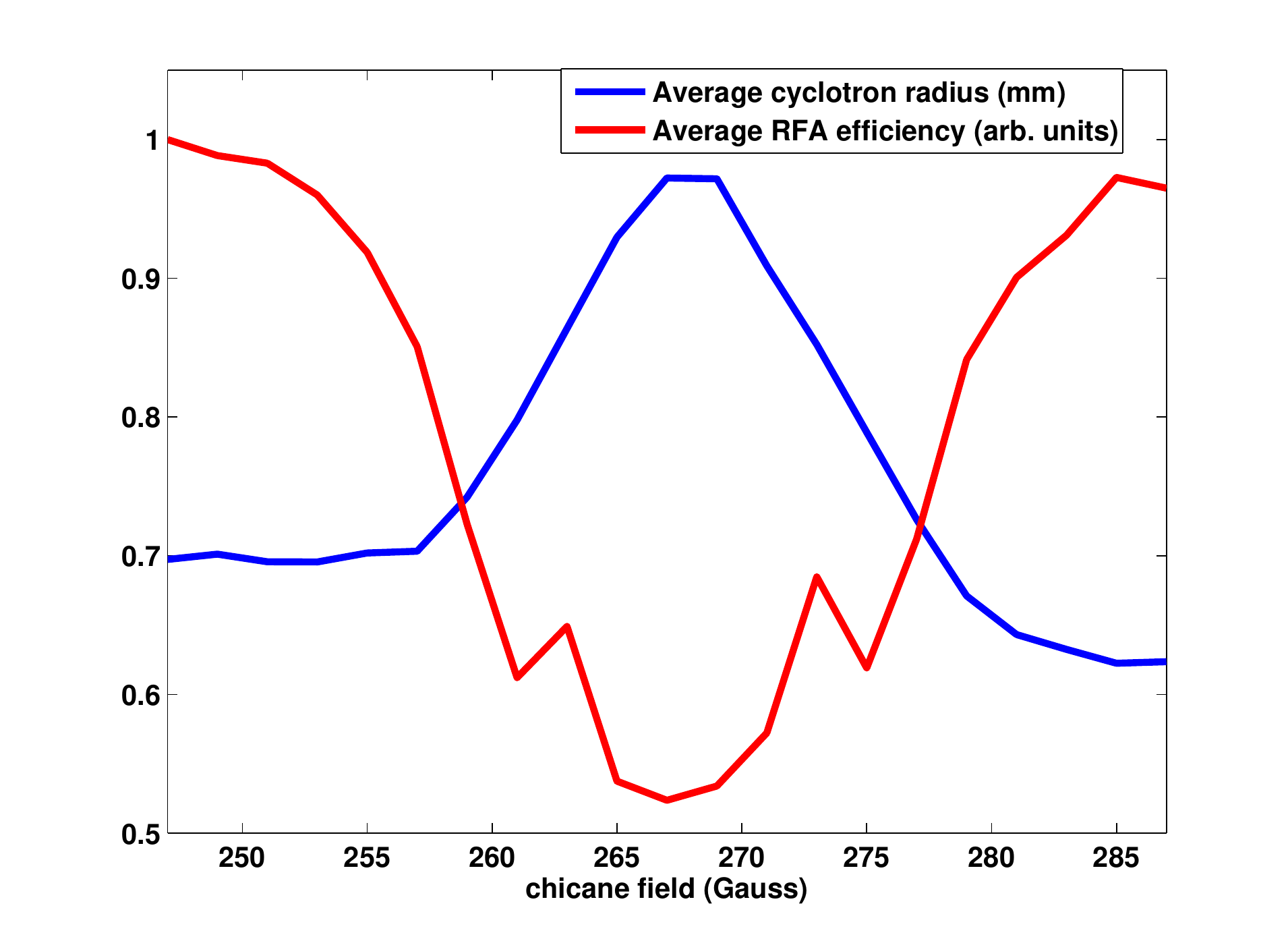} \\ 
\caption[Effect of cyclotron resonance on RFA efficiency]{\label{fig:cyc_eff} Effect of cyclotron resonance on RFA efficiency, 1x45x1~mA $e^+$, 4~ns, 5~GeV.  Under the resonant field, the average electron cyclotron radius increases, resulting in a decrease in the average RFA efficiency.}
\end{figure}

Under the conditions of a cyclotron resonance,  we expect to see a increase in the RFA signal, due to the increased energy of the cloud electrons.  As discussed in Section~\ref{ssec:cyc_res}, we do indeed observe peaks in the RFA current in the aluminum chicane chamber, but in the TiN-coated chambers we observe dips.  Fig.~\ref{fig:cyc_sim} shows a simulated magnetic field scan over a cyclotron resonance, in both an aluminum and TiN-coated chamber.  Consistent with the data, we observe an increase in the aluminum chamber signal, but a decrease in the TiN chamber signal.  Fig.~\ref{fig:cyc_eff} provides an explanation: since the additional energy in the resonant electrons comes from transverse beam kicks, these electrons will have a larger cyclotron radius, and thus a lower RFA efficiency (see Fig.~\ref{fig:dipole_rfa_eff}).  Thus there are two competing effects: an increased cloud density due to a higher average SEY, and lower overall detector sensitivity.  In the aluminum chamber (where the peak SEY is high) the former effect dominates, while in the coated chamber (where the peak SEY is low) the latter one does.  The net result is resonant peaks in the uncoated chamber, and dips in the coated one.

\subsection{Simulation of Anomalous Enhancement in the Wiggler RFA}

The main disadvantage of treating the RFA analytically (as described in Section~\ref{sec:modeling}) is that we cannot self-consistently model any interaction between the detector and the cloud, such as the trampoline effect described in Section~\ref{ssec:tramp}.  Motivated by these measurements, we have incorporated into POSINST a model of the RFA geared toward reproducing the geometry of the RFAs installed in the wiggler vacuum chambers.  The motion of the electrons within the RFA, including the electrostatic force from the retarding field, is tracked using a special add-on routine.  The grid is modeled realistically, and secondary electrons can be produced there, with the same secondary yield model used for normal vacuum chamber collisions.  The peak secondary electron yield and peak yield energy can be specified separately for the grid.  Because the actual retarding field is included in the wiggler RFA model, the retarding voltage must be specified in the input file, and a separate simulation must be run for each voltage.

\begin{figure}
   \centering
   \includegraphics*[width=0.6\textwidth]{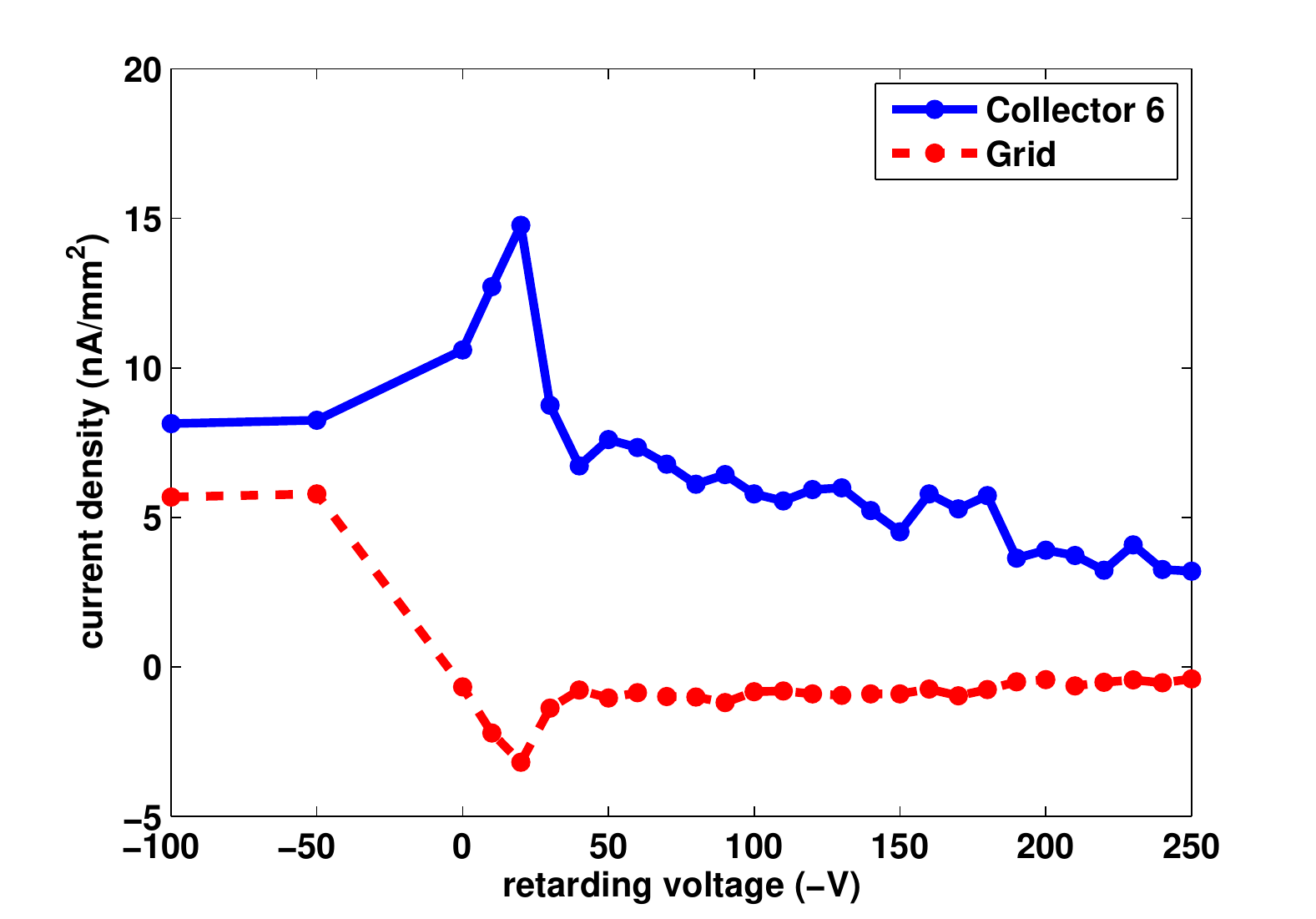} 
   \caption[POSINST simulation showing resonant enhancement]{\label{fig:int_model} POSINST simulation showing resonant enhancement in a wiggler RFA, 1x45x1.2~mA $e^+$, 2.1~GeV, 14~ns, central collector.  Compare to Fig.~\ref{fig:tramp_example}.}
\end{figure}

Fig~\ref{fig:int_model} shows the result of running this full particle tracking simulation, for the set of beam conditions corresponding to Fig.~\ref{fig:tramp_example}.  Notably, the simulation reproduces the resonant enhancement seen in the data, at approximately the same voltage ($\sim$10~V for 14~ns spacing), and shows that the extra signal comes from the grid.


\section{Conclusions}

Electron cloud buildup has been investigated in dipole field regions throughout CESR.  Measurements of multipacting and cyclotron resonances have been made at different bunch spacings, bunch currents, and with electron and positron beams.


A sophisticated analytical model for multipacting resonances has been developed, which takes into account secondary emission energy, as well as the time for kicked electrons to reach the chamber wall.  This model is generally consistent with data, and has been further validated by computer simulations.  An anomalous enhancement in the center-pole wiggler RFA signal has also been identified as an artificial multipacting resonance.

Cyclotron resonances have been observed in the chicane RFAs, at field values that correspond well to basic theory.  The question of these resonances sometimes appearing as dips, rather than peaks in the signal, has been explained as a detector efficiency effect.




The electron cloud density is very sensitive to multipacting effects.  On resonance, we observe as much as a factor of 3 increase in electron cloud signal for positron beams, and several orders of magnitude for electron beams (though the measured signal for electron beams was always lower than for positrons).  Because electron cloud is a potential limiting factor for high current, low emittance beams, avoiding these resonances is crucial for achieving emittance and stability goals in present and future accelerators.

\begin{acknowledgments}

This research was supported by NSF and DOE Contracts No. PHY-0734867, No. PHY-1002467, No. PHYS-1068662, No. DE-FC02-08ER41538, No. DE-SC0006505, and the Japan/U.S. Cooperation Program.


The authors would like to thank D. Rubin, G. Dugan, J.A. Crittenden, J. Sikora, J. Livesey, M. Palmer, and K. Harkay for their helpful advice and suggestions; R. Schwartz, S. Santos, and S. Roy for assisting with the RFA measurements; and M. Furman at LBNL for his support with the POSINST simulation code.

\end{acknowledgments}


\bibliographystyle{medium}
\bibliography{rfa_field_prst}

\end{document}